# Manifestation of central symmetry of the celestial sphere in the mutual disposition and luminosity of the Quasars

Iurii Kudriavtcev

*We performed the check of supposition about the possibility of manifestation of the previously observed phenomenon of central symmetry of the celestial sphere through existence of the opposite quasars. We discovered the existence of some pairs of quasars located opposite each other with close by form profiles magnitudes of luminosity in the ranges u, g, r, i, z, when correlation coefficient close to 1. We discovered that the percentage of the pairs with correlation coefficients Rxy>0.98 for the opposite located quasars is significantly higher than that for the random pairs.*

*The analysis of the dependence of this exceedance from the artificial breaking of the central symmetry has shown, that it practically disappears with symmetry breaking by more than 0.05 degrees. Thus we can confirmed the manifestation of the central symmetry of celestial sphere through existence of the central symmetrical pairs of quasars, which can be interpreted as the pairs of images of the same object.*

*We shown the possibility of a theoretical modeling of the observed dependencies in the closed Universe model. It can be supposed, that a relatively small amount of the discovered pairs of the opposite quasars is conditioned by the fact, that the opposite objects for most of the quasars are galaxies, which are not included to the chosen as initial source of data quasar catalog SDSS-DR5.*

**98.80.-k**

## 1. Introduction

Central symmetry of the celestial sphere that was observed earlier in the phenomena of the central symmetry and central antisymmetry of the microwave [1], can also be manifested relatively to the mutual disposition of the distant observed objects. In this relation we analyze in this work the data about the most distant observed objects – quasars, for the purpose to establish the presence or the absence of the characteristics of the central symmetry in their disposition that is provoked by the possibility of the simultaneous observation of the signals that have gotten to us from the same source by two different arcs of the big circle of the closed Universe. For the observer these signals will seem to be the ones from two different sources located in the two opposite (centrally symmetrical) points of the celestial sphere.

To perform this analysis we used the materials from SDSS Quasar Catalog IV, DR5, that contains the data about 77429 quasars [2].

## 2. Selection of the pairs of the opposite and neighboring quasars.

To compare the coordinates of the opposite quasars we can divide all the objects in the catalog to two groups that are located in the opposite hemispheres, i.e. at the values of the right ascension from $0^0$ to $180^0$ and from $180^0$ to $360^0$. The quasars of the first groups will be the opposite ones in relation to the quasars of the second group that satisfy to the following conditions for the right ascension (RA) and declension (DE):

$$RA_1 = RA_2 - 180^0; \quad (1)$$

$$DE_1 = -DE_2; \quad (2)$$

That is why to find the pairs of the opposite quasars the coordinates of the quasars of the first group should be compared to the inverted coordinates of the second group

$$RA_{2inv} = RA_2 - 180^0; \quad (3)$$

$$DE_{2inv} = -DE_2. \quad (4)$$

To reach it we made a general list of the quasars from the catalog that contains the quasars from the first group with the true coordinates $RA_1$ и $DE_1$, and the quasars from the second group with the inverted coordinates $RA_{2inv}$ и $DE_{2inv}$, and created a map of their location. It appeared that because of the inhomogeneous distribution of the visible quasars the greatest amount of the quasars with the close values $RA_1$ and $RA_{2inv}$, $DE_1$ and $DE_{2inv}$, lay within the narrow subequatorial zone of the sky ($-1.25^0 < DE < +1.25^0$). That is why the future analysis was made withing the zone ($-1.4^0 < DE < +1.4^0$), that include 12759 quasars, including 7183 of the quasars from the first group (RA from $0^0$ to $180^0$) and 5576 quasars from the second group (RA from $180^0$ to $360^0$). The reason why we chose a wider zone will be explained below.

In the chosen zone we selected the pairs of quasars whose mutual angular distance is less then the distance to any other quasar that is not a part of the pair. As a result of this procedure we selected 3888 pairs of quasars, including 1327 pairs of quasars that belong to the different celestial hemispheres (opposite ones), and 2561 pairs of quasars that belong to the same celestial hemisphere (neighboring ones). We will call them respectively "pairs 2" and "pairs 1".

**3. Analysis of the mutual disposition.**

On the picture 1 we show the distribution of "pairs 2" by the angular distance R between the members of the pair. $R=((RA_1-RA_{2inv})^2+(DE_1-DE_{2inv})^2)^{1/2}$. Points on the histogram correspond to the quantity of the pairs with the value of R within the interval between the current and the following point. The first point corresponds to $0<R<0.01$ of degree.

Strict centrally symmetrical disposition of the quasars in the pair would correspond to R=0, but from the Picture 1 we see that this variant does not practically realize, and for most of the pairs R is within the range from 0.02 to 0.15 of a degree with the maximum about 0.05 - 0.07 of a degree. Because of that and to avoid loosing the information the width of the studied zone of the celestial sphere was increased as it was mentioned above from ±1.25 to ±1.4 of a degree.

To discuss the mode of the demonstrated distribution and its reasons let us move to the pairs of the neighboring quasars ("pairs 1").

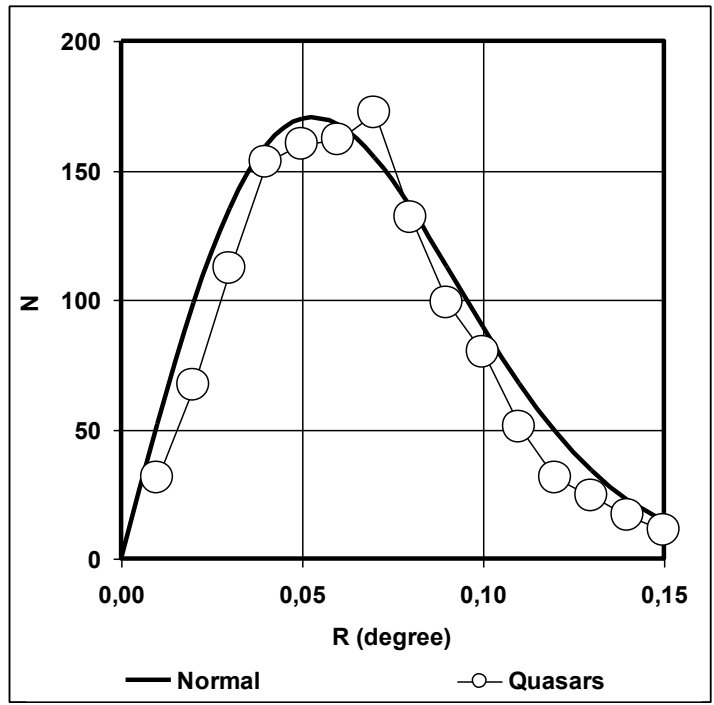

**Picture 1.** Distribution of the pairs of the opposite quasars "pairs 2" by the angular distance R between the members of the pair (deviation from the strictly opposite position). The points on the histogram correspond to the quantity of the pairs with the values of R within the range between the current and the following point. в интервале между данной и предыдущей точкой. The solid line is the distribution that correspond to the normal distribution of the deviation of the coordinates (see discussion below).

It is obvious that the amount of "pairs 1" includes the pairs formed as the result of the known effect of the gravitational lensing, i.e. deviation of the rays by the gravitational fields on their way so the observer sees not only one but several (two or more) images of the initial source. As for making a picture of the same object we must keep the condition of the approximate equality of the redshift, they can be selected from the basic mass by the characteristic of the small difference of the redshifts $dZ = |Z1-Z2| \ll 1$.

Supposing that the most of the pairs with the small difference of the redshifts were formed as a result of the gravitational lensing and that the influence of the gravitational fields on the dispersing of the images should increase with the increasing of the distance to the radiation source (Z), let us see their distribution in the coordinates (R,Z).

On the Picture 2 we show the distributions R(Z) for the different scopes of the selection and values dZ. $Z = (Z1+Z2)/2$.

a), b) — for the first quarter of the analyzed zone $(0^0<RA_1<45^0),(180^0<RA_2<225^0)$,

c), d) - for the whole analyzed zone $(0^0<RA_1<180^0),(180^0<RA_2<360^0)$.

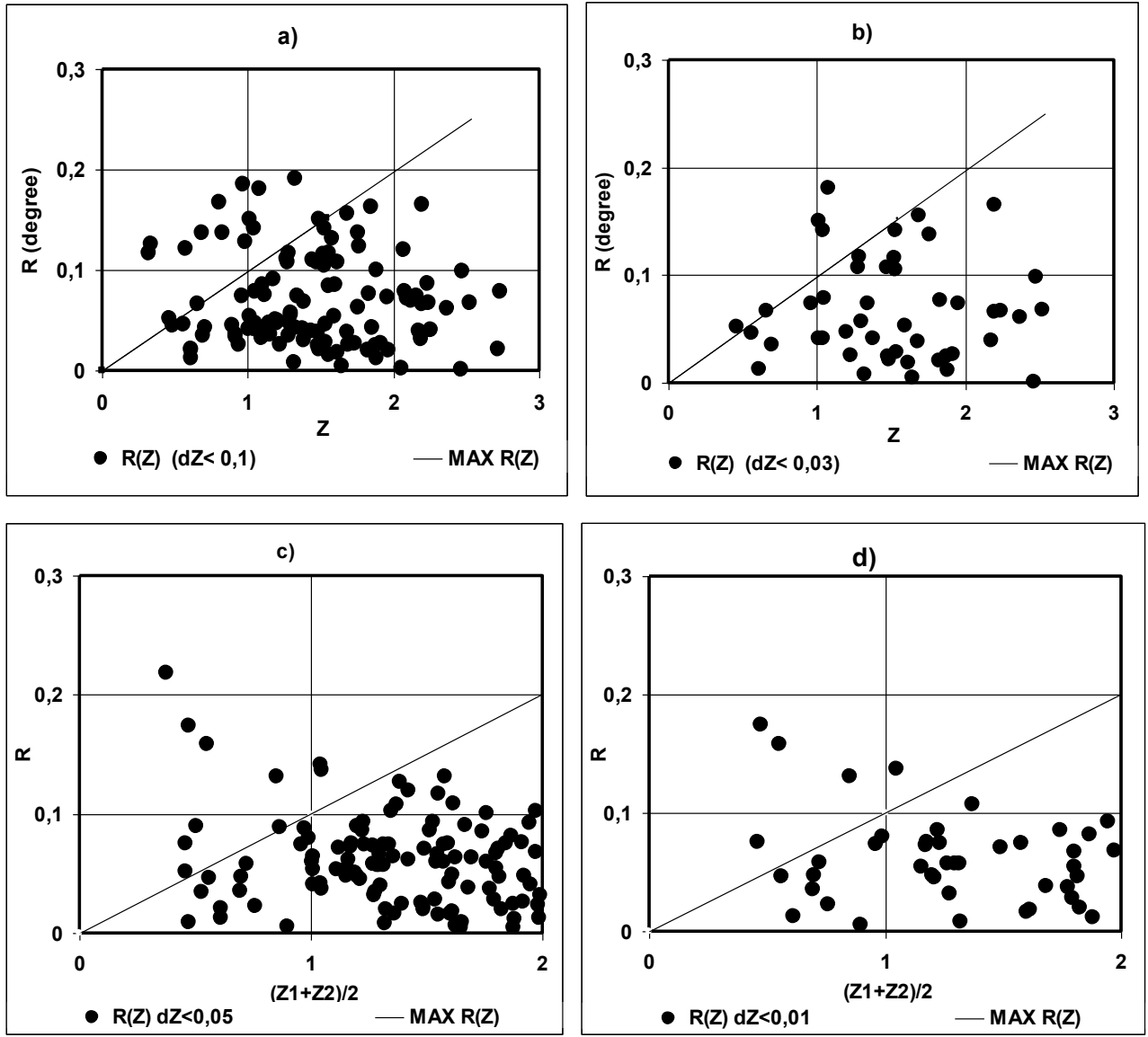

**Picture 2.** Distribution of the "pairs 1" in coordinates (R,Z) for different selection scopes and values dZ. The straight lines mark the limits of R(Z), below which the basic mass of the pairs that were supposedly formed as a result of the gravitational lensing.

On all the distributions of Picture 2 the most of the pairs with the small values of dZ are compactly located below the lines of R=0.1Z that limit the area of the location. It can be considered as a circumstantial proof of their formation as a result of the gravitational lensing and dispersing of the rays while moving through the space where the effect of (R) should increase with the increasing of the distance to the radiating object (Z). Supposing that this dependence is close to the linear one we will build up the dependence of the quantity of pairs from the relation (R/Z), which, in this case, is the quantitative characteristic of the object image dispersing that appeared as a result of the gravitational lensing depending from the distance to it. According to the Picture 2, the basic mass of these pairs should be located within the range 0<(R/Z)<0.1.

The obtained distributions are shown on the Picture 3. The same way as the distribution on the Picture 1, they have the maximum lying at R ≈ (0.04-0.05) of a degree, but are characterized by bigger influence of a random component in consequence of a small amount of the analyzed pairs that were selected by the criterion of smallness dZ.

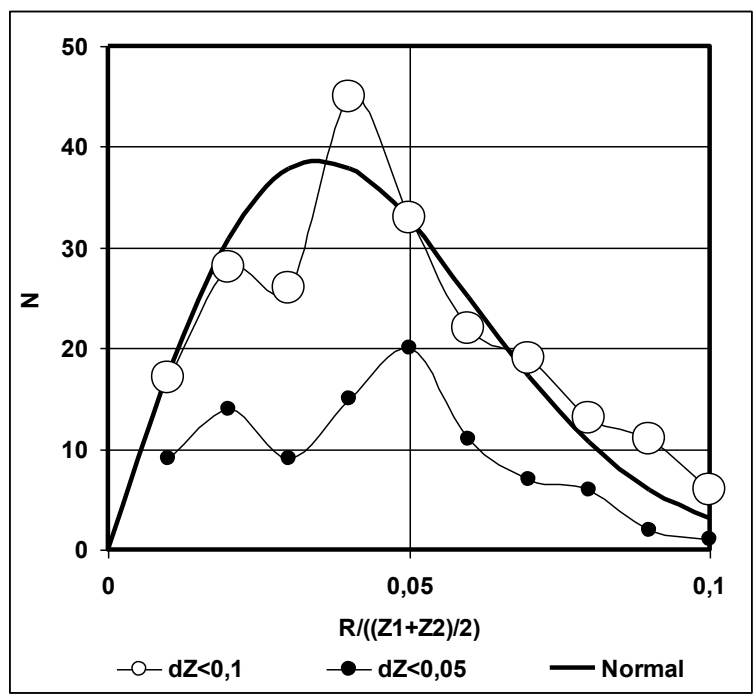

**Picture 3.** Distribution of the pairs consisting of the quasars from the same celestial hemisphere "pairs 1" that were selected by the criterion of the small difference of the redshifts dZ, by R/Z. The solid line is the distribution that corresponds to the normal distribution of the deviation of the coordinates (see the discussion below).

The radiation of the distant objects when moving to the observer crosses a lot of gravitational fields of random orientation. That is why the direction of the resulting dispersing of their images whose value is characterized on the Picture 3 by the value r =R/Z, can be considered random for any range dr evenly distributed on the surface of the ring S= $2\pi r$ dr. At that the density of the distribution of the dispersed images by r will be proportional to N/r.

Dependence $\rho = N/r$ from r for the pairs, corresponding to the curve of Picture 3 (dZ<0.1) is shown on the Picture 4. We see that it can be approximately described by the normal distribution

$$\rho = K\exp(r^2/2\sigma^2). \qquad (12)$$

The solid line shows the normal distribution with the dispersion $\sigma_1 = 0.035$. The dependence that corresponds to this normal distribution

$$N(r) \sim \rho r = Kr \exp(r^2/2\sigma^2). \qquad (13)$$

is shown by the solid curve on the Picture 3.

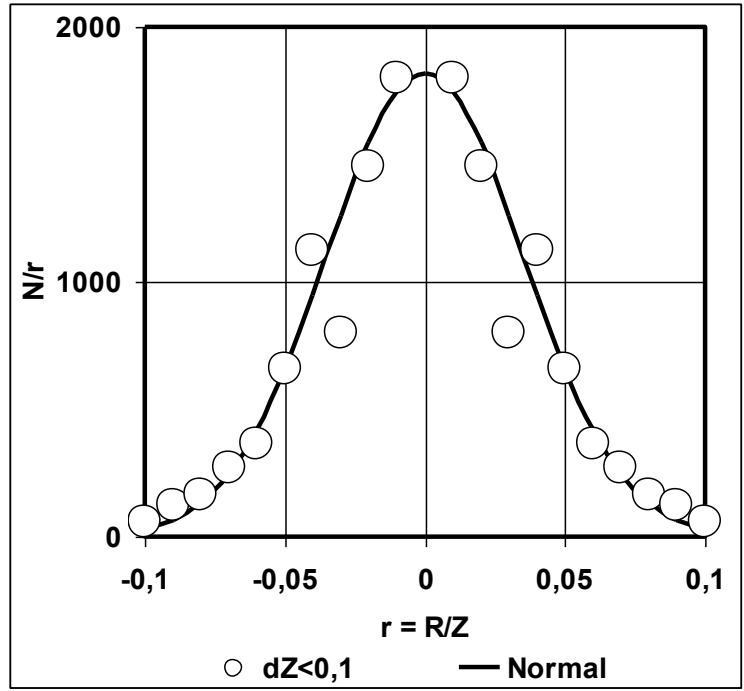

**Picture 4.** Distribution of the deviations (dispersion) of the images of the quasars due to the gravitational lensing according to the information of the Picture 3. The solid line is the normal distribution at σ = 0.035.

Getting back to the pairs of the opposite quasars we can note that if the opposite images ("pairs 2") are also the images of the same source that reached the observer by two different ways (by different arcs of the big circle), the effects of the gravitational dispersion should influence them the same way as it influences the described above "pairs 1".

In this case the dependence N(R) with the consideration of the supposition about the linear mode of R(Z) will be defined by the expression

$$N(R) \sim \rho r = Kr \exp(r^2/2\sigma_2^2); \qquad (13)$$

where $\sigma_2 = Z \sigma_1$, $Z = (Z_1+Z_2)/2$.

The solid line on the Picture 1 shows the dependence of N(R), calculated by the expression (13) for

$$\sigma_2 = 0.053 \approx 1.5 \sigma_1. \qquad (14)$$

This way the distribution N(R) of the pairs of the opposite quasars shown on the Picture 1 is described accurately enough by the expression for the effect of the gravitational dispersion of the images of the same source at $Z = (Z_1+Z_2)/2 \approx 1,5$. Here $Z_1$ and $Z_2$ are the redshifts of the source when observed by two parts of the big circle.

On the Picture 5 we show the distribution of "pairs 2" by $Z=(Z_1+Z_2)/2$. The distribution has a delineated maximum. The average value $Z_{mean}=1.49\pm0.42$, which corresponds to the correlation $\sigma_2/\sigma_1$ (14).

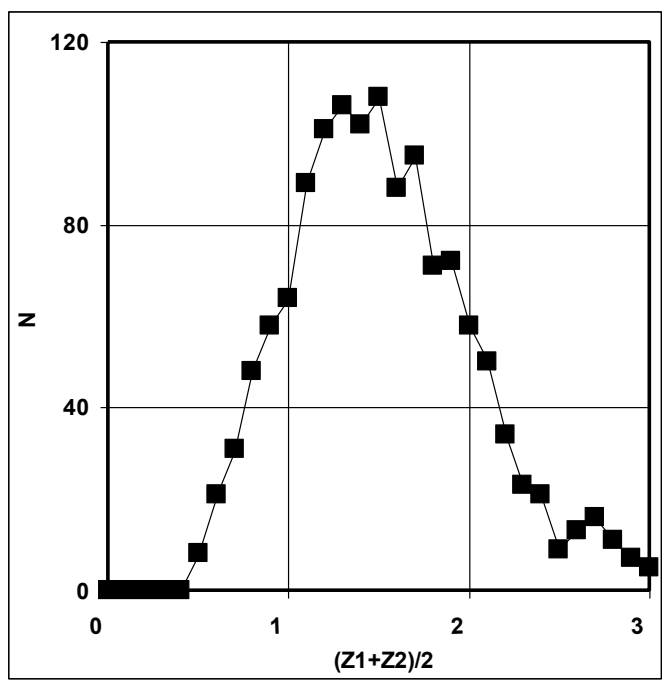

**Picture 5.** Distribution of the pairs of the opposite quasars by the values of the average redshift $Z_m = (Z_1+Z_2)/2$.

## 4. Comparative analysis of the luminosity profiles

The described particularities of the mutual disposition do not allow identify with confidence the detected pairs of the opposite quasars as the pairs of the images of the same object. For this we should make an additional comparative analysis by spectral or other individual characteristics of the observed objects. The most accessible but not the most accurate way to analyze is to compare the profiles of the magnitudes of the luminosity of quasars in the ranges **u, g, r, i, z** (from 300 to 1000 nm)**,** which data can be found in the catalog [2] for almost all of the quasars.

N the Picture 6 we show the forms of these profiles for several randomly chosen quasars from the first dozen (by entry number, «recno») of the catalog [2]. Despite the fact that all the shown profiles has the mode, descending from «**u**» to «**z**», the forms of the profiles of different quasars significantly differ which allows to hope to the possibility of their use for identification.

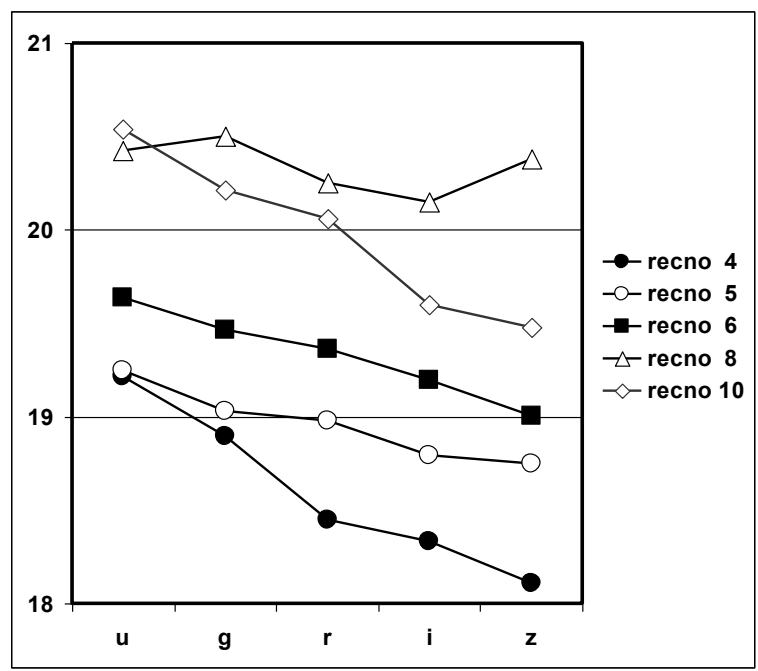

**Picture 6.** Different forms of the profiles of the magnitude of the quasar luminosity.

Let us check the possibility of use of this identification method on the example of the pairs of neighboring quasars ("pairs 1") some of which can be images of the same object that were split as a result of the gravitational lensing.

On the Picture 7 we show the example of the comparison of the profiles for the pairs of neighboring quasars with angular distance R<0.1 of degree and small difference of redshifts which allow to consider them with high probability the pairs of the images of the same object. For each pair we calculated the value of the Pearson correlation coefficient $R_{xy}$. Сочетание существенного сходства форм профилей и высоких значений коэффициента корреляции подтверждают, что каждая из этих пар квазаров с большой вероятностью является парой изображений одного и того же объекта. Таким образом, метод сопоставления профилей может быть использован для идентификации объектов.

Let us try to apply it to the pairs of the opposite quasars. On Pictures 8, 9 we show the fragments of the map of celestial sphere with the pairs of opposite quasars located there. Light circles show the location of the quasars in true coordinates (RA,DE), dark ones — quasars with inverted coordinates ($RA_{inv}$,$DE_{inv}$), transferred from the opposite points of the celestial sphere (members of "pairs 2"), light circle of bigger size – quasars that are the closest to the inverted ones (members of "pairs 2"). On the histograms under the fragments of the maps we show the luminosity magnitudes profiles for the members of the pairs and other closely located quasars. Corresponding correlation coefficients are in the Table 1.

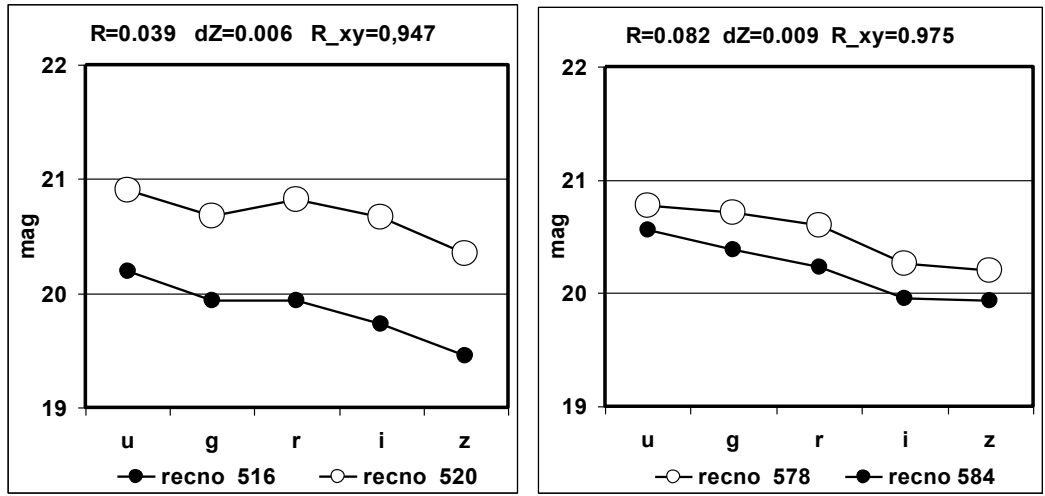

**Picture 7.** Example of comparison of the profiles of the pairs of neighboring quasars with the small difference of redshifts.

For all the fragments shown on the Pictures 8,9 we can note that the profiles of the opposite quasars – members of the "pairs 2" – are closer to each other by form than to the neighboring quasars and the corresponding coefficients of the correlation are closer to 1.

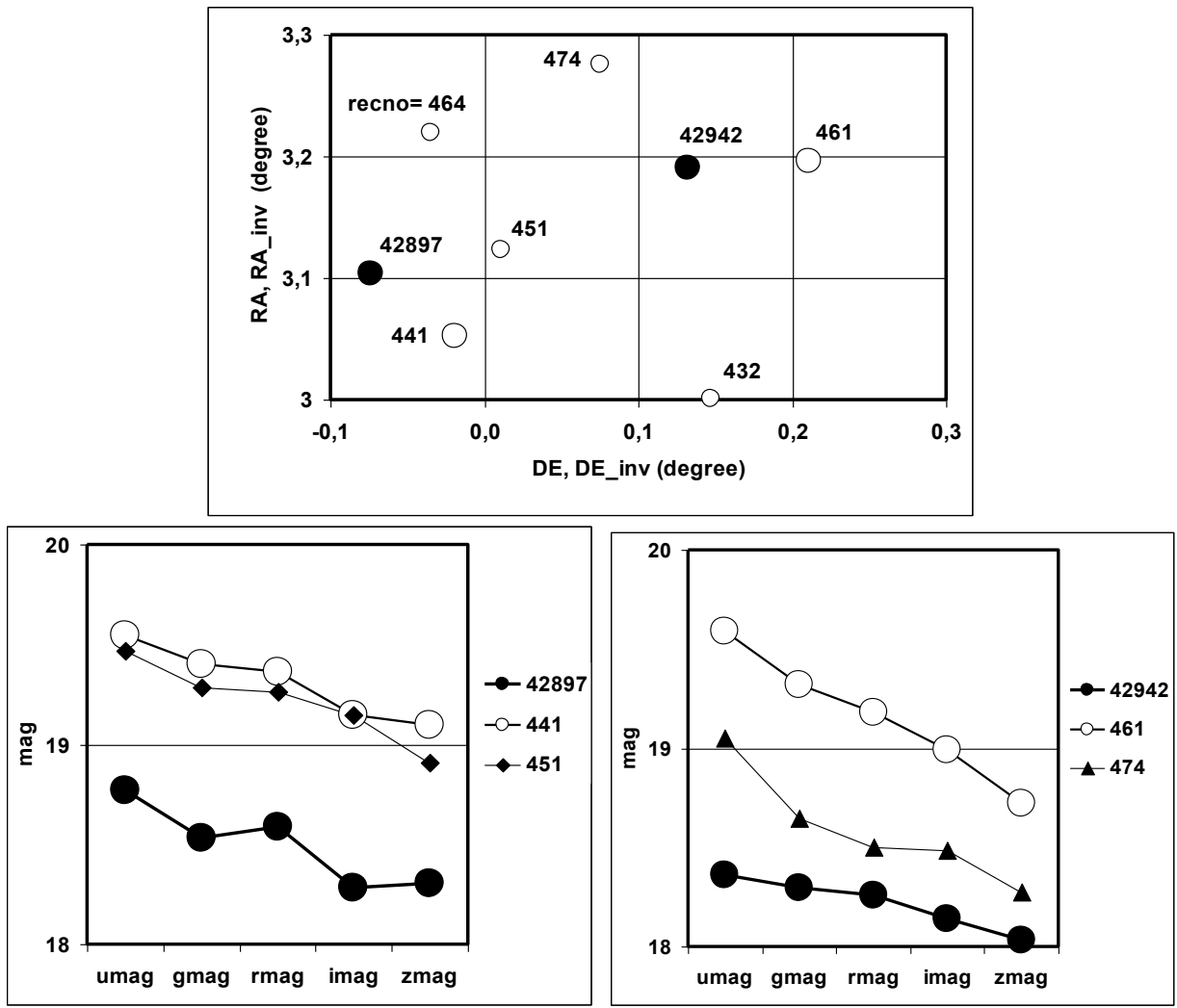

**Picture 8.** Fragment of the map of the celestial sphere with two pairs of the opposite quasars and the histograms of their luminosity magnitudes profiles. Light circles – quasars in true coordinates (RA,DE), dark ones — quasars with inverted coordinates (RA$_{inv}$,DE$_{inv}$), transferred from the opposite points of the celestial sphere.

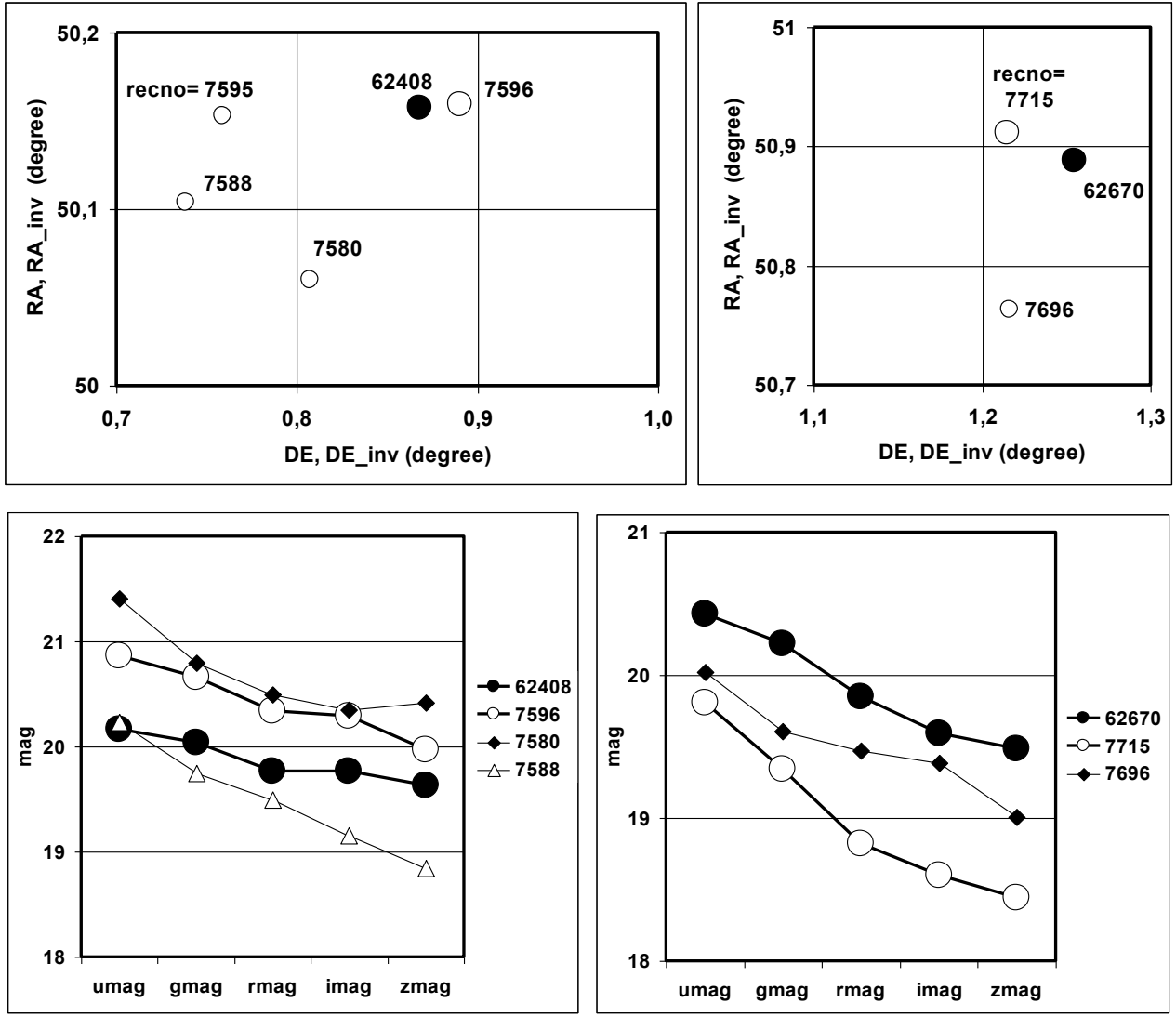

**Picture 9.** Two fragments of the map of the celestial sphere with the pairs of the opposite quasars and histograms of the luminosity magnitudes profiles.

| № | recno 1 | recno 2 | $R_{xy}$ (1,2) |
|---|---|---|---|
| 1 | **441** | 451 | 0,943 |
| 2 |  | **42897** | **0,976** |
| 3 | **461** | 474 | 0,952 |
| 4 |  | **42942** | **0,982** |
| 5 | **7596** | 7580 | 0,877 |
| 6 |  | 7588 | 0,978 |
| 7 |  | **62408** | **0,986** |
| 8 | **7715** | 7696 | 0,944 |
| 9 |  | **62670** | **0,989** |

**Table 1.** Pearson correlation coefficients corresponding to the pairs of quasar profiles shown on the Pictures 8,9. The registration numbers of the opposite quasars – members of the "pairs 2" – and corresponding correlation coefficients are marked bold.

On the Picture 10 we show several more examples of the luminosity magnitudes profiles of the pairs of opposite quasars with high correlation coefficients ($R_{xy}>0.95$). Numerical characteristics of these pairs are in the Table 2.

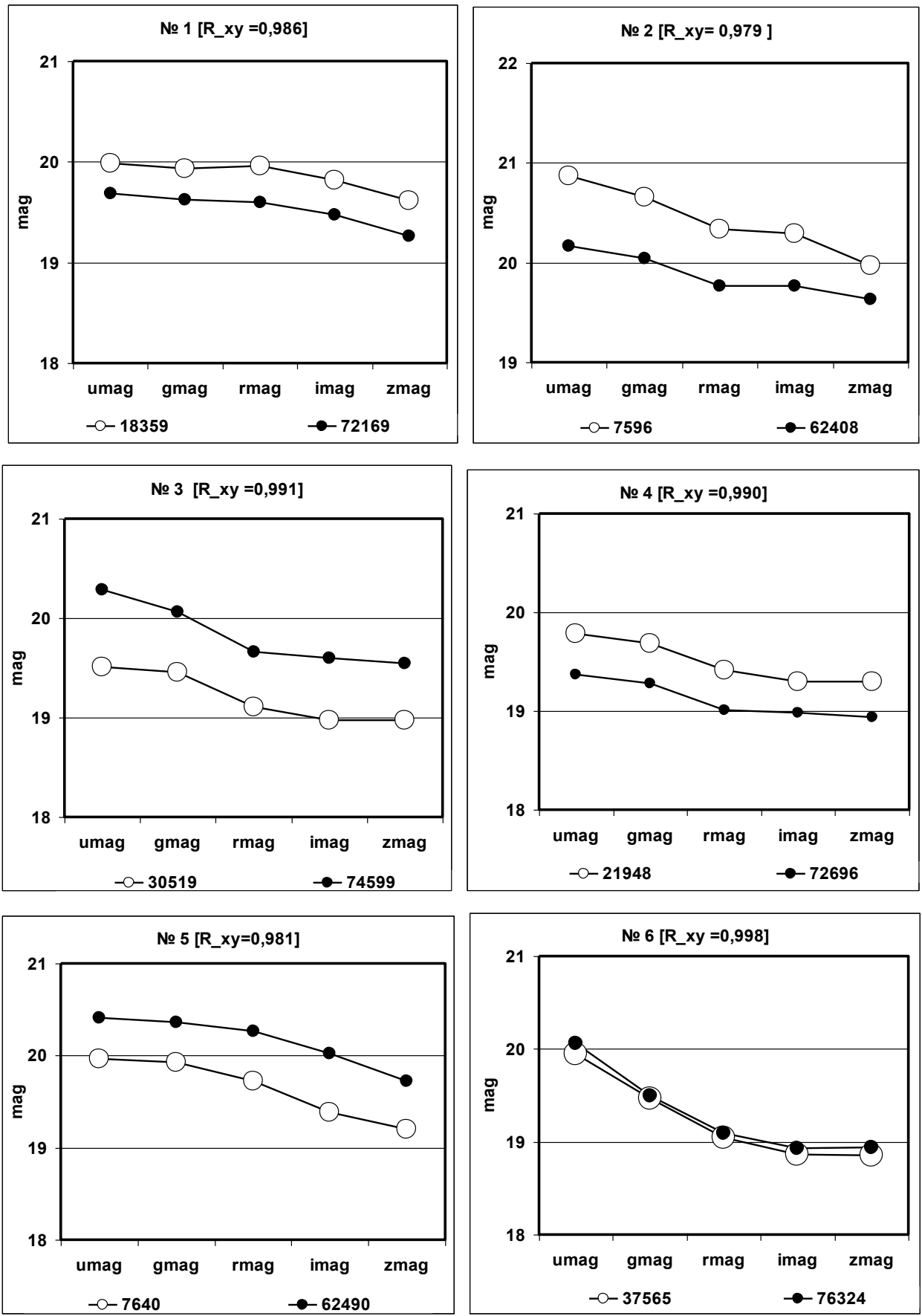

**Picture 10.** Several examples of the comparison of the luminosity profiles of the pairs of opposite quasars with high correlation coefficients ($R_{xy} > 0.95$).

| № | recno 1 ($0^0$<RA<$180^0$) | recno 2 ($180^0$<RA<$360^0$) | R (degree) | |dZ| | (Z1+Z2)/2 | R_xy (1,2) |
|---|---|---|---|---|---|---|
| 1. | 7596 | 62408 | 0,022 | 0,098 | 1,13 | 0,986 |
| 2. | 7640 | 62490 | 0,029 | 0,041 | 2,01 | 0,979 |
| 3. | 18359 | 72169 | 0,070 | 0,109 | 1,96 | 0,991 |
| 4. | 21948 | 72696 | 0,042 | 0,109 | 1,28 | 0,990 |
| 5. | 30519 | 74599 | 0,041 | 0,110 | 1,33 | 0,981 |
| 6. | 37565 | 76324 | 0,059 | 0,165 | 1,29 | 0,998 |

**Table 2.** Numerical characteristics of the pair of opposite quasars shown on the Picture 10.

## 5. Statistical analysis of the correlation coefficient of luminosity profiles.

The brought data on similarity of luminosity profiles of opposite quasars confirms the supposition that some of the opposite pairs of quasars can be the pairs of images of objects that came to us by different arcs of the closed Universe. However they cannot be considered the proof of existence of the mentioned effect because the closeness of angular coordinates and luminosity magnitudes profiles can be the result of the random coincidences chosen from a big amount of the catalog objects (77429) [2].

To check the randomness or regularity of the discovered particularities we made a statistical analysis of the distributions of correlation coefficients of luminosity magnitudes profiles of the pairs of opposite quasars. Distribution of Pearson correlation coefficient for 1327 detected "pairs 2" was compared to the similar distribution for the multitude of the random pairs of quasars. For this 12759 quasars of the analyzed zone of the celestial sphere ($-1.4^0 <$ DE $< +1.4^0$) were combined in random pairs by assigning to them some random numbers and than sorting out. As a result we formed a set of 40000 randomly made pairs. Analysis of the distribution of correlation coefficient made on this multitude has shown that it is significantly not symmetrical in relation to the area of definition (-1,1) and depends on the difference of redshifts of the members of pairs, that are shifting in the direction of $R_{xy} = 1$ at decreasing dZ. Here and further under dZ we understand the absolute value (dZ ≡ |dZ|).

On the Picture 11 we show the distributions by $R_{xy}$ for the whole set of 40000 random pairs and for the subsets of the chosen pairs with restrictions of dZ.

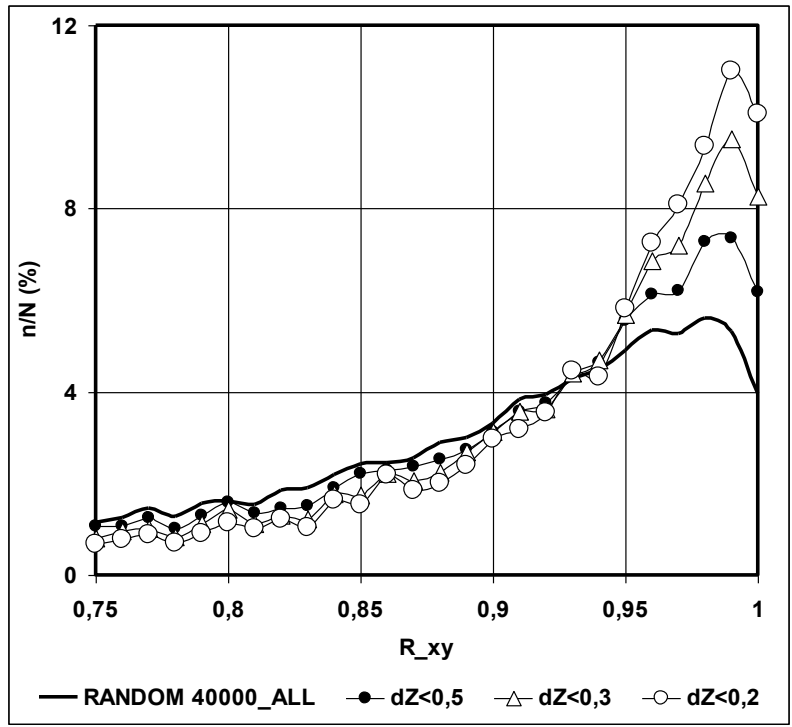

**Picture 11.** Distributions by $R_{xy}$ of luminosity magnitudes profiles of 40000 randomly formed pairs of quasars and pairs with dZ<0.2, dZ<0.3, dZ<0.5 chosen from this set.

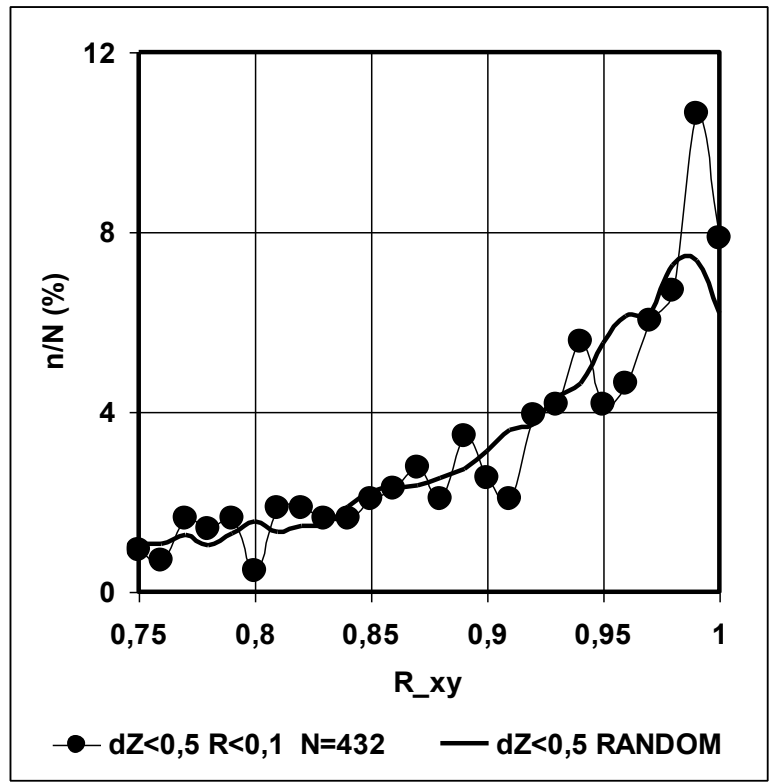

**Picture 12.** Distribution by $R_{xy}$ of the luminosity profiles of the pairs of opposite quasars with $dZ<0.5$ (selection scope — 432 pairs) in comparison with the distribution for the random pairs with $dZ<0.5$.

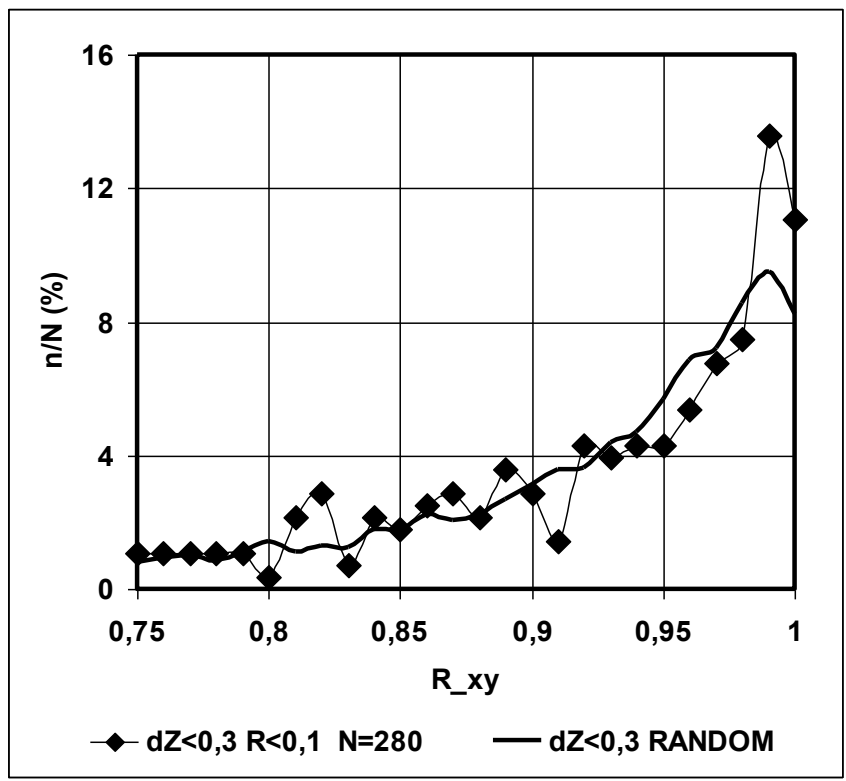

**Picture 13.** Distribution by $R_{xy}$ of the luminosity profiles of the pairs of opposite quasars with $dZ<0.3$ (selection scope — 280 pairs) in comparison with the distribution for the random pairs with $dZ<0.3$.

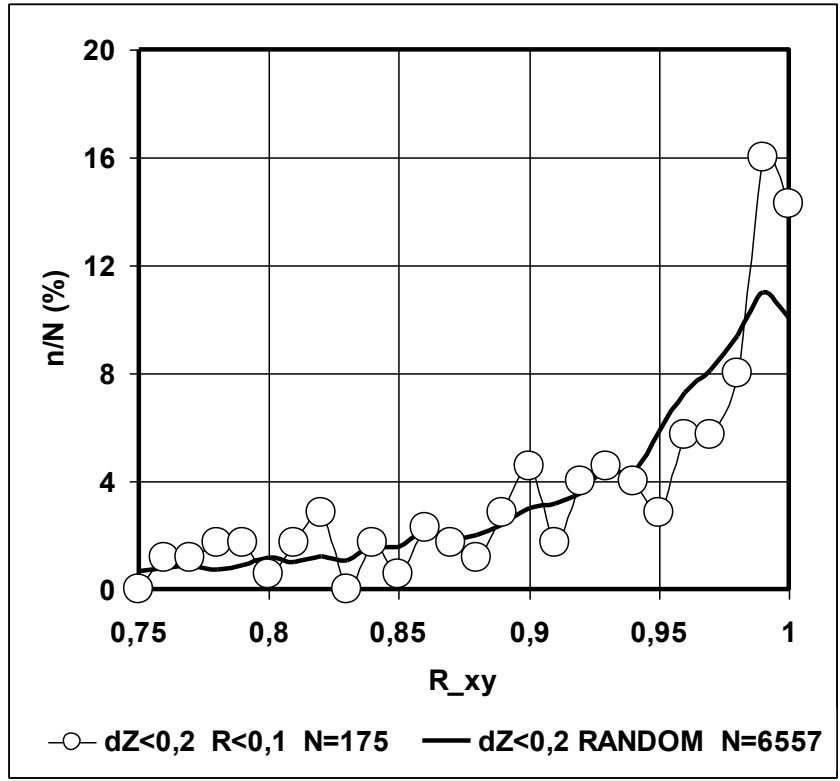

**Picture 14.** Distribution by $R_{xy}$ of luminosity profiles of the pairs of opposite quasars with dZ<0.2 (selection scope – 175 pairs) in comparison with the distribution for random pairs with dZ<0.2.

As it is evident from the Pictures 12-14, for the opposite quasars the fraction of the pairs with the biggest values of the correlation coefficient (two last points in the distribution corresponding to 0.98>$R_{xy}$>0.99 and 0.99>$R_{xy}$>1) appears to be much bigger than for the random pairs with the same range dZ. However the general number of the opposite pairs with such big values of $R_{xy}$ appears to be comparatively not that big which still allows us to interpret the obtained data as not related with the possibility to observe the signals that are coming from the same source from two different directions but as random ones.

In this relation we made an additional check of the obtained data. It included formation and selection of the pairs of opposite quasars with artificial deviation from the condition of the central symmetry ($RA_{inv}$ = RA-$180^0$; $DE_{inv}$ = -DE). This deviation was defined by adding the shift Shift_$RA_{inv}$, where $RA_{inv}$ was defined by the expression

$$RA_{inv} = RA - 180^0 + Shift\_RA_{inv}; \qquad (15)$$

and using the same method we performed the procedures of selection and analysis of the opposite pairs for different values of Shift_$RA_{inv}$ in the range from -$1^0$ до +$1^0$ (-$1^0$, -$0.5^0$, -$0.2^0$, -$0.1^0$, -$0.05^0$, **$0^0$**, $0.05^0$, $0.1^0$, $0.2^0$, $0.5^0$, $1^0$). For each of these values Shift_$RA_{inv}$ we built up the dependencies построены n/N($R_{xy}$), similar to the ones shown on the Pictures 12-14.

On Pictures 15-17 we show the final dependencies from Shift_$RA_{inv}$ of containing in these distributions of the amount of opposite pairs with bigger $R_{xy}$ in comparison with the distributions for the random pairs, i.e. differences (n/N(%)$_{quas}$-n/N(%)$_{rand}$).

Picture 15 — for $R_{xy} > 0.97$ (sum of differences $(n/N(\%)_{quas} - n/N(\%)_{rand})$ for three last points of the curves similar to the ones shown on the Pictures 12-14 for Shift_RA$_{inv}$ = 0).

Picture 16 — for $R_{xy} > 0.98$ (by two last points of the curves).

Picture 17 — for $R_{xy} > 0.99$ (by last points of the curves).

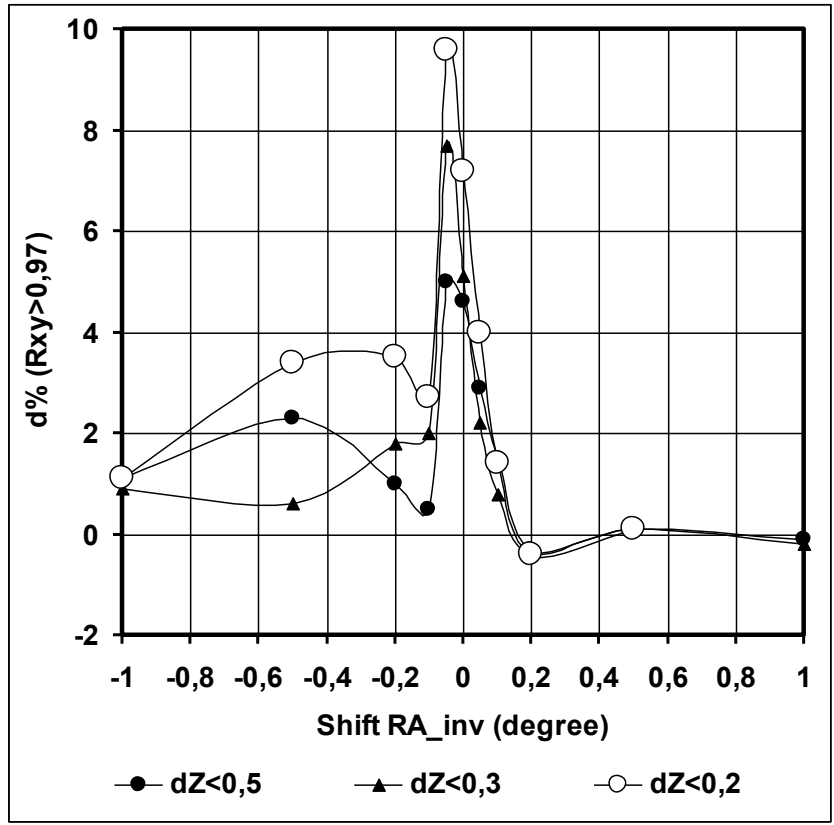

**Picture 15.** Dependencies from Shift_RA$_{inv}$ of the percentage of the opposite pairs of quasars with bigger $R_{xy}$ in comparison with distributions for the random pairs (for dZ<0.5, dZ<0.3, dZ<0.2, **$R_{xy}$ >0.97**).

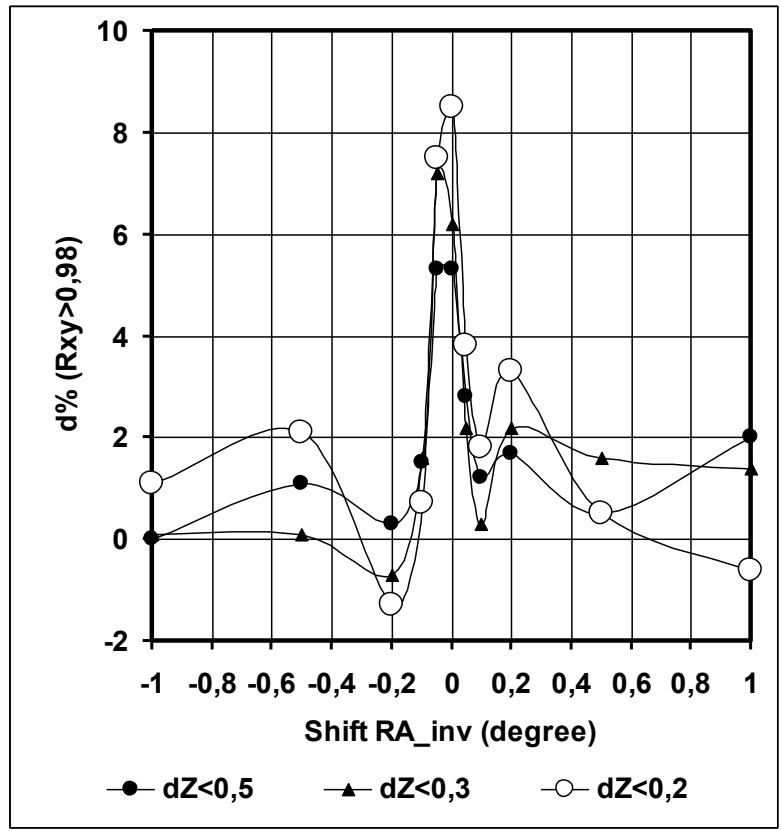

**Picture 16.** Dependencies from Shift_RA$_{inv}$ the percentage of the opposite pairs of quasars with big R$_{xy}$ in comparison with the distributions for the random pairs (for dZ<0.5, dZ<0.3, dZ<0.2, **R$_{xy}$ >0.98**).

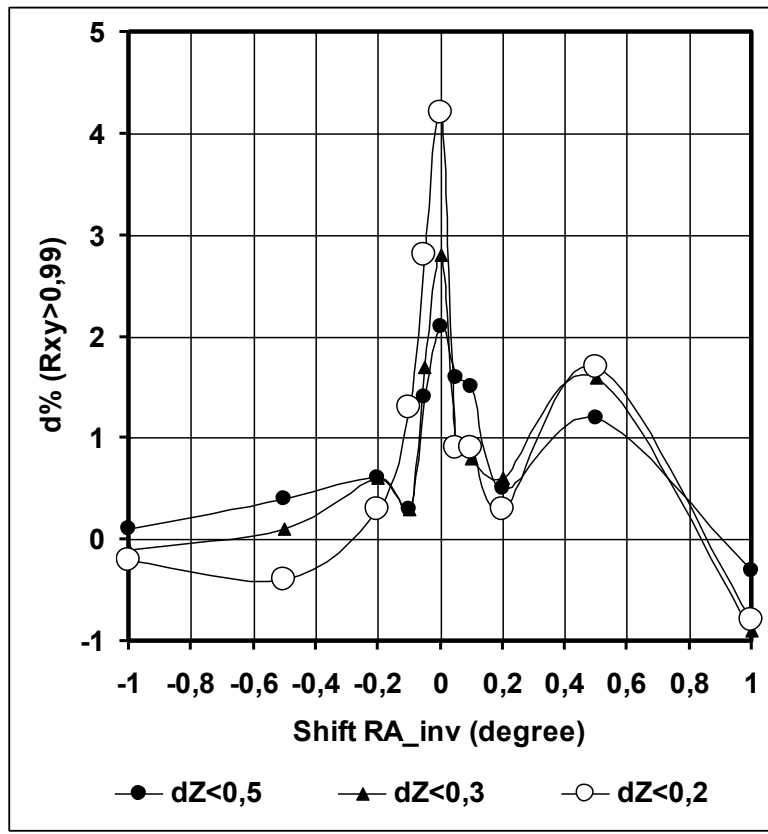

**Picture 17.** Dependencies from Shift_RA$_{inv}$ the percentage of the opposite pairs of quasars with big R$_{xy}$ in comparison with the distributions for the random pairs (for dZ<0.5, dZ<0.3, dZ<0.2, **R$_{xy}$ >0.99**).

The results demonstrated on Pictures 15-17 show that the significant difference in percentage of the pairs with the highest R$_{xy}$ between the pairs of opposite and random quasars is observed only at Shift_RA$_{inv}$ ≈ 0, i.e. only for the really opposite quasars. This way they can be considered the proof of the existence of central symmetry related to the possibility to observe the images of the same object that came to the observer by the different arcs of the closed Universe as it is impossible to suppose a random appearance of the particularities shown on the Pictures 15-17, because of the utmost small probability of such an event. Let us note that this particularity has not a local but a general nature as the discovered pairs are almost homogeneously spread in the ranges of the angles of the analyzed zone of the celestial sphere where there are the observed quasars, see Picture 18.

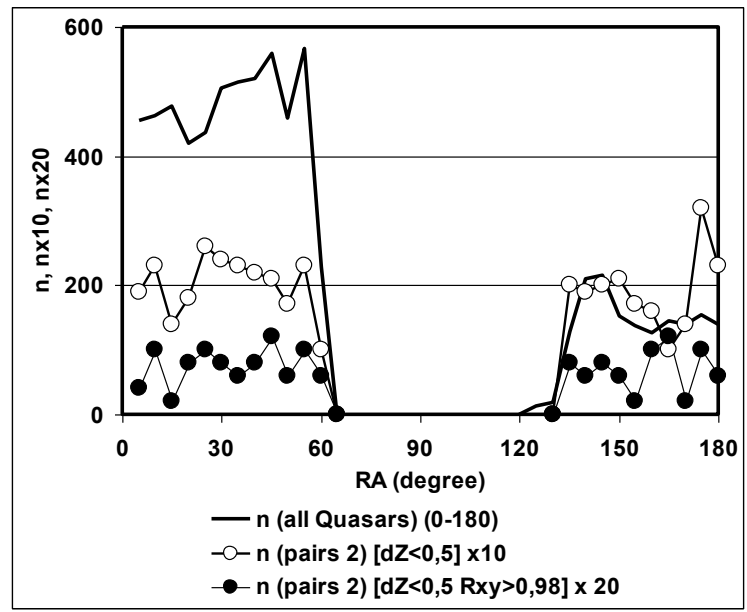

**Picture 18.** Homogeneity of the spatial distribution of the discovered pairs of the opposite quasars.

Probability of the entering of the random pair of quasars to the range $R_{xy}$ >0.98 is, according to the data of the Picture 11 (sum of the last two points on the curves) 13% for dZ<0.5, 17% for dZ<0.3, 21% for dZ<0.2. Quantities of the superfluous pairs with $R_{xy}$ >0.98 by the data of Pictures 12-14 and Picture 16 are: 22 pairs for dZ<0.5, 20 pairs for dZ<0.3, 16 pairs for dZ<0.2. Probability of the random run shown on the Picture 16, is $(0,13)^{22} \approx 3*10^{-20}$ for dZ<0.5, $4*10^{-16}$ for dZ<0.3, $1*10^{-11}$ for dZ<0.2.

There is a question about the small amount of the discovered pairs of opposite quasars which can be identified by high values of luminosity profiles correlation coefficients as a pair of images of the same object as in case of existing of the phenomenon it should be related to all of the observed objects the same way.

One of the reasons – chosen in this work simple method of selection of the initial pairs by mutually minimal angular distance $R = ((RA-RA_{inv})^2+(DE-DE_{inv})^2)^{1/2}$, where all the pairs that are randomly located near any other quasar are excluded. It is rather difficult to evaluate the loss at the moment.

Another reason – chosen method of the analysis of the individual characteristics of the objects by luminosity magnitudes profiles in 5 ranges that does not allow to consider the shift of the signals frequencies because of the redshift. It limited our analysis by pairs of quasars with small difference of redshifts that make a rather small part of the general amount of the selected pairs, i.e.   432 / 1327 ≈ 33% for dZ<0.5,  21% for dZ<0.3, 13%  for dZ<0.2.

One more reason that can significantly limit the amount of the observed centrally symmetrical pairs will be discussed in the following section.

## 6.  Distribution of the quasars and pairs of opposite quasars by redshift and possible approached to the theoretical modeling of these distributions in the closed Universe.

In the closed Universe that is viewed as a 3-dimensional hypersphere with the radius "a" that expands in a 4-dimensional space, the velocity of the movement of the observed object relatively to the observer is defined by the velocity of stretching (increasing of the radius) of the hypersphere and by angular distance between the object and the observer.  Graphic image of this model in a 2-dimensional section is shown on the Picture 19.

If the radius ща the hypersphere increases with the relative velocity $\beta_0 = V/c = da/cdt$, the relative velocity of the object moving-off from the observer when the object is on the angular distance $\varphi$, and without considering the relativistic effects will be equal to $\beta_0\varphi$, and the difference of the relative velocities of the moving-off from the observer objects that are on angular distances $\varphi$ and $(\varphi+d\varphi)$, will be defined by the expression $d\beta=\beta_0 d\varphi$. Dependence of the velocity of moving-off from $\varphi$ with the consideration  of the relativistic effects will be defined from the law of  composition of velocities:

$$v = (v' + V) / (1 + v'V/c^2); \qquad (16)$$

where v — velocity of the object in the stationary frame of the observer, V — velocity of the object in the frame that moves relatively to the observer along the ray of vision with the velocity v',  c — light speed. If the object  located on the angular distance of  $\varphi$,  is moving off from the observer with the velocity v', and another object located by $d\varphi$ farther, is moving off from the first with the speed V = $c\beta_0 d\varphi$, the velocity of the second object relatively to the observer will be defined by the expression (16).

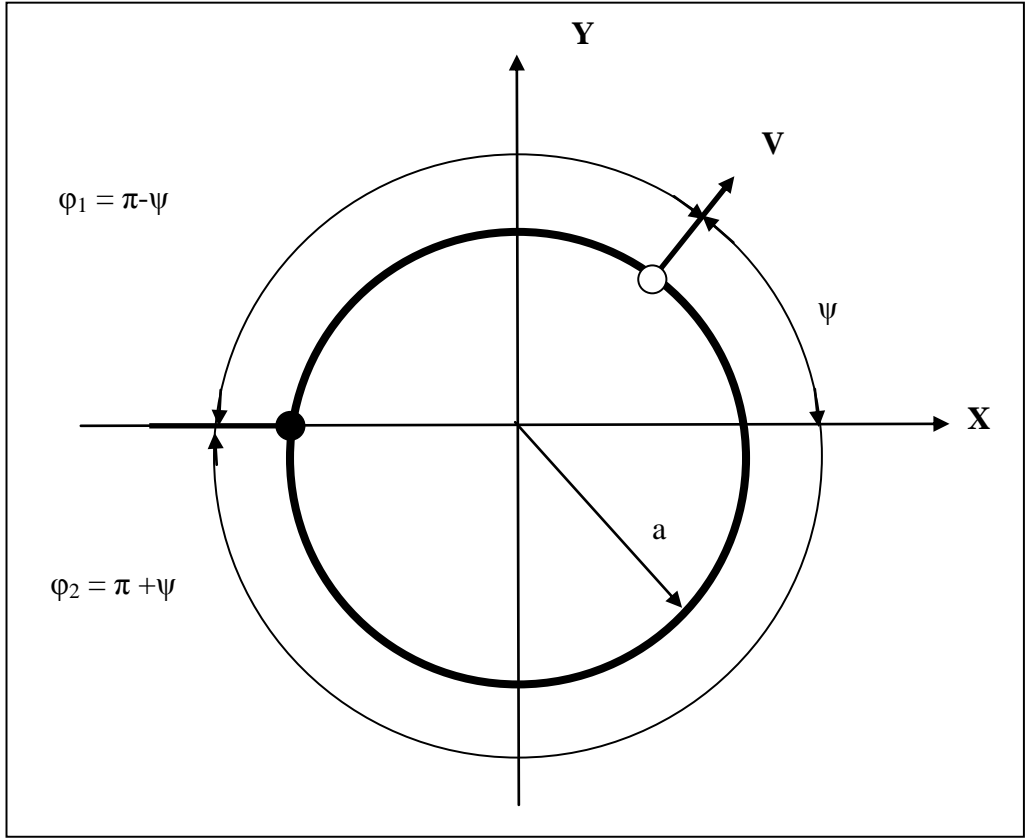

**Picture 19.** Two-dimensional section of the 4-dimensional space with the hypersphere of radius "a". Angular coordinates $\varphi_1$ and $\varphi_2$ characterize the distance from the radiating object (white spot) to the observer (black spot) by two parts of the big circle.

Considering the velocity V small (at small $d\varphi$  V<<v'<c, and consequently v'V<<$c^2$), we can transform (16) the following way:

$$v = (v' + V) / (1 + v'V/c^2) \approx (v' + V)(1 - v'V/c^2) =$$
$$= v' + V(1 - (v'^2 + v'V)/c^2) \approx v' + V(1 - v'^2/c^2); \qquad (17)$$

From where

$$\Delta v = v - v' = V(1 - v'^2/c^2); \qquad (18)$$

Adding the relative velocity $\beta = v/c$, and turning to the differentials we obtain:

$$d\beta = \beta_0 d\varphi \, (1 - \beta^2); \qquad (19)$$

Taking integral we obtain definitely

$$\int d\beta/(1 - \beta^2) = \text{arth}(\beta) = \beta_0 \int d\varphi = \beta_0 \varphi; \qquad (20)$$

from where

$$\beta(\varphi) = \text{th}(\beta_0 \varphi); \qquad (21)$$

redshift Z accordingly to the Doppler effect will be related to $\beta$ and $\beta_0$ by the expressions

$$Z = ((1+\beta)/(1-\beta))^{1/2} - 1 = \exp(\beta_0 \varphi) - 1. \qquad (22)$$

Average value of the redshifts $Z_m = (Z_1 + Z_2)/2$ of two object images being observed on the angular distances $\varphi_1$ and $\varphi_2$, will be expressed via the angle $\psi = (\varphi_2 - \varphi_1)/2$ by formula

$$Z_m = \exp(\beta_0 \pi) \text{ch}(\beta_0 \psi) - 1. \qquad (23)$$

Let us see within this model the distribution of the pairs of the opposite quasars by the average redshift $Z_m$, shown on Picture 5.

According to the requirement of homogeneity of the Universe we consider the observed objects (quasars) to be homogeneously distributed in the space i.e. their spatial density $\rho = N/V$ is constant., where N - general amount of the objects in the closed Universe, V - volume of its space, equal to $V = 2\pi^2 a^3$ [3]. From that

$$\rho = N/2\pi^2 a^3. \qquad (24)$$

The amount of the objects located within the distance "l" in the layer with the width "dl", will be equal to

$$n(l, dl) = \rho S dl; \qquad (25)$$

where distance "l" is related to the angular coordinate $\varphi$ by the expression $l = \varphi a$, and S - cross-sectional area of the space that depends on the angular coordinate as

$$S = = 4\pi a^2 \sin^2 \varphi. \qquad (26)$$

From that

$$n(\varphi, d\varphi) = (2N/\pi) \sin^2 \varphi \, d\varphi; \qquad (27)$$

Let us relate the angular coordinate of the object with its redshift Z.

From (22)

$$\beta_0 \varphi = \ln(Z+1); \qquad (28)$$

$$dZ/d\varphi = \beta_0 \exp(\beta_0 \varphi); \qquad (29)$$

from where the amount of objects in the layer with the redshift Z with the width dZ

$$n(Z) = (K/\beta_0) \sin^2[\ln(Z+1)/\beta_0] \exp(-\beta_0 \varphi); \qquad (30)$$

where $K = \text{const} = (2N/\pi) dZ$.

Considering (28) we will obtain definitely

$$n(Z) = (K/((Z+1)\beta_0)) \sin^2[\ln(Z+1)/\beta_0]; \qquad (31)$$

Now let us see the distribution of the pairs of the opposite images of quasars that are located on the angular distances $\varphi_1 = \pi - \psi$ and $\varphi_2 = \pi + \psi$ (Picture 19). The amount of the pairs of objects' images is proportional to the amount of the objects themselves $n(Z)$. As the images are located on the different distances that correspond to the different redshifts we will study their distribution by $Z_m = (Z_1 + Z_2)/2$. From (28) we will obtain

$$Z_m = \exp(\beta_0 \pi) \operatorname{ch}(\beta_0 \psi) - 1. \qquad (32)$$

Let us express $n(\varphi, d\varphi)$ in (27) via $\psi$, considering that $\sin(\psi) = \sin(\varphi_1) = \sin(\varphi_2)$:

$$n(\psi, d\psi) = (2N/\pi) \sin^2 \psi \, d\psi; \qquad (33)$$

and relate $\psi$ с $Z_m$. From (32)

$$(Z_m + 1) = \exp(\beta_0 \pi) \operatorname{ch}(\beta_0 \psi); \qquad (34)$$

$$d(Z_m + 1)/d\psi = \beta_0 \exp(\beta_0 \pi) \operatorname{sh}(\beta_0 \psi); \qquad (35)$$

$$\beta_0 \psi = \operatorname{arch}[(Z_m + 1)/\exp(\beta_0 \pi)]; \qquad (36)$$

$$d\psi = d(Z_m+1)/(\beta_0 \exp(\beta_0\pi) \, sh(\beta_0\psi)) = d(Z_m+1)/(\beta_0 \exp(\beta_0\pi) \, (ch^2(\beta_0\psi)-1)^{1/2}). \qquad (37)$$

Let us denote $x = ch(\beta_0\psi)$. Из (34)

$$x = ch(\beta_0\psi) = (Z_m+1)/\exp(\beta_0\pi). \qquad (38)$$

Then

$$d\psi = dx/(\beta_0(x^2-1)^{1/2}); \qquad (39)$$

$$\psi = (1/\beta_0) \, arch \, x. \qquad (40)$$

Substituting (39),(40) to (33) and considering that $dx = d(Z_m+1)/\exp(\beta_0\pi)$ from (38), we will obtain $n(Z_m)$:

$$n(Z_m) = K \sin^2[arch \, x \,/\beta_0](x^2-1)^{-1/2} \beta_0^{-1} \exp(-\beta_0\pi). \qquad (41)$$

Here $K = const = (2N/\pi)dZ_m$, $x = (Z_m+1)\exp(-\beta_0\pi)$.

When calculating by (41) it is necessary to consider that $[arch x /\beta_0] = \psi$ ($0<\psi<\pi$), i.e. $Z_m$ in (41) has physical meaning only within the limits that correspond to $0<\psi<\pi$.

On Picture 20 we show the calculated by (41) dependence $n(Z_m)$ for different values of $\beta_0$, on Picture 21 we compare the distribution of the pairs of opposite quasars (Picture 5) with the calculated distribution for $\beta_0 = 0.25$.

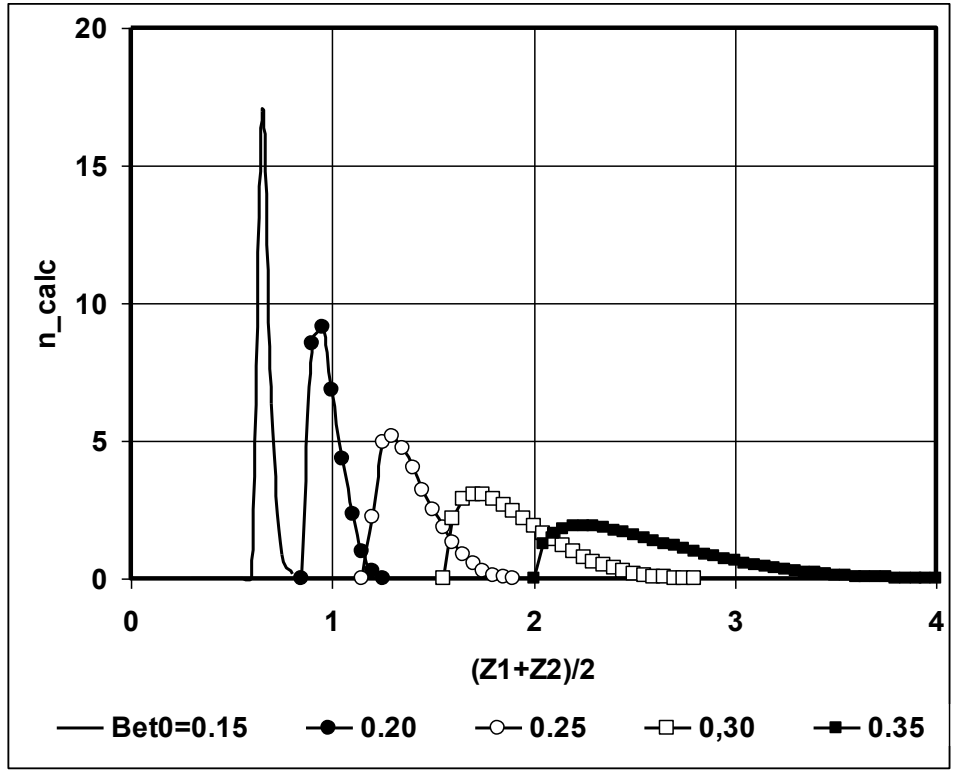

**Picture 20.** Calculated by the expression (41) dependencies $n(Z_m)$ for different $\beta_0$.

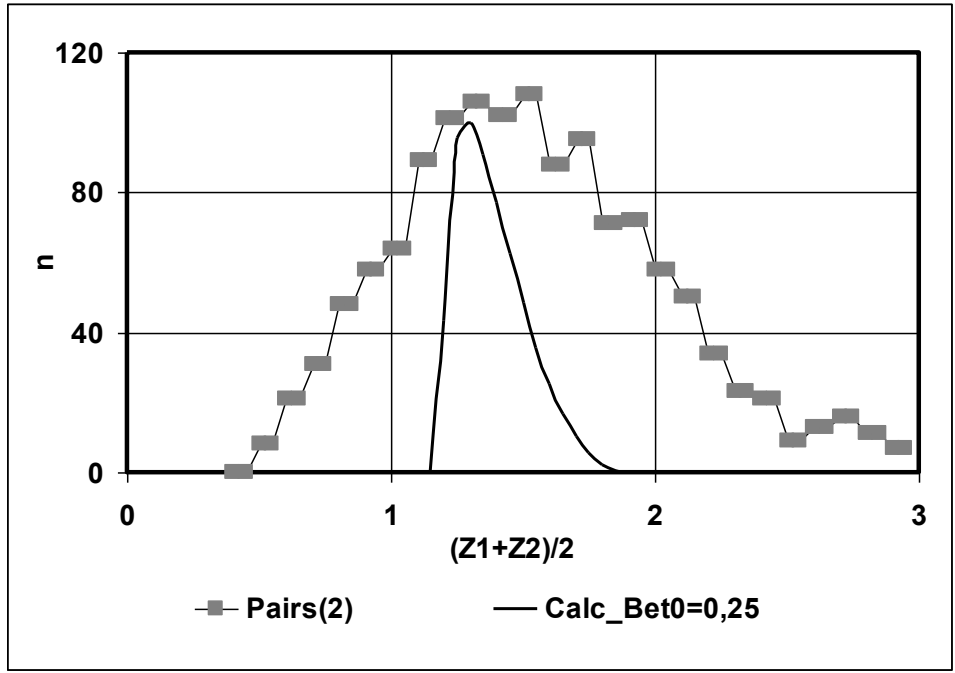

**Picture 21.** Distribution by $Z_m = (Z1+Z2)/2$ of the pairs of opposite quasars (Picture 5) in comparison with the dependence $n(Z_m)$, calculated by the expression (41) for $\beta_0 = 0.25$.

Distribution of the pairs of opposite quasars has a significantly bigger width than the calculated curve $n(Z_m)$. The nature of the change of the curves for different $\beta_0$ (Picture 20) allows to suppose that such a widening in relation to the theoretical curse can appear if the values $\beta_0$ in the different sections of the space are not the same. In this case different pairs of quasars whose signals come through the areas of the space with different average values of $\beta_0$, will be described by the different dependencies $n(Z_m)$, and the real distribution will represent an integral curve that is wider than any of the curves $n(Z_m)$ separately.

This supposition does not contradict to the existing data about the structure of the Universe where the congregations of the galaxies with high density of the matter are separated by the empty spaces. At that the radius of the curvature of the space should change the way that it should represent not a smooth hyperspheric surface but rather some wrinkly "hyper-newspaper" where all the fringes and brows in the places where the matter is congregated are separated by flat sections of equal size and form. As the rays of light that came to us from the different angles of the celestial sphere came by different ways on this "hyper-newspaper" i.e. through the sections of the space with different values of the curvature then the average values of $\beta_0$ can also be different.

To check it we made calculations supposing that the distribution of the values of $\beta_0$ has the nature of the "normal distribution" with average value of 0.25, for the values of the standard deviation from 0.02 to 0.06. As at these values the normal distribution lies in the area $0.1 < \beta_0 < 0.4$ (Picture 22), when calculating we summarized the dependencies $n(Z_m)$ for $\beta_0$ in the range from 0.1 to 0.4 with pitch 0.01, multiplied on the weight coefficients of the normal distributions, Picture 22.

The results of the calculations are shown on the Picture 23 in comparison with the curve for $\beta_0 = 0.25$.

On Picture 24 we show the calculated curve $n(Z_m)$ for normal distribution $\beta_0$ with standard deviation 0.06 and corrected average value 0.26 in comparison with the distribution of the pairs of opposite quasars shown on the Picture 5.

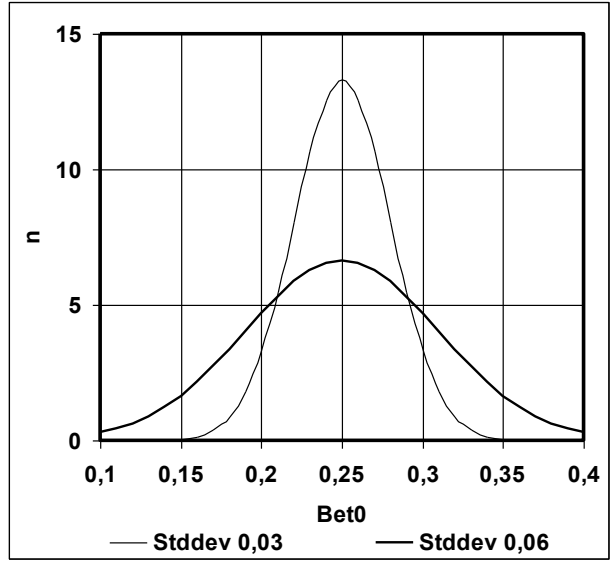

**Picture 22.** Normal distributions used at the calculation of the dependencies $n(Z_m)$, shown on the Picture 23.

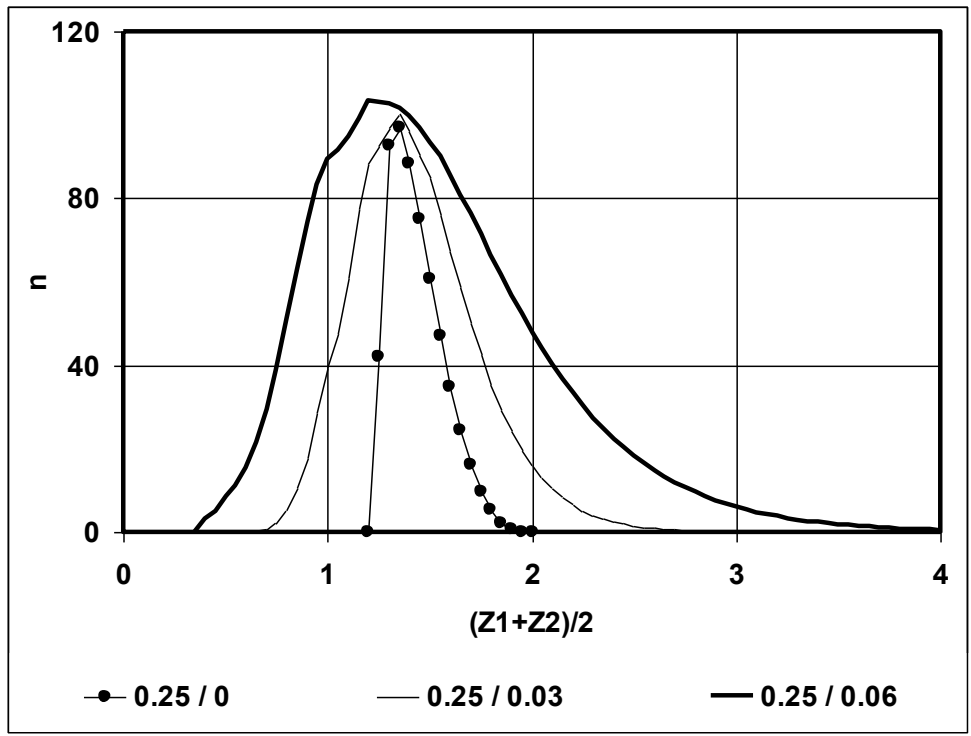

**Picture 23.** Calculated distributions of the pairs of the opposite quasars by value of the average redshift $Z_m$ at constant $\beta_0 = 0.25$ and at $\beta_0$ accordingly to the distributions of Picture 22. Smoothed to compensate the inhomogeneities provoked by the discrete nature of the calculation.

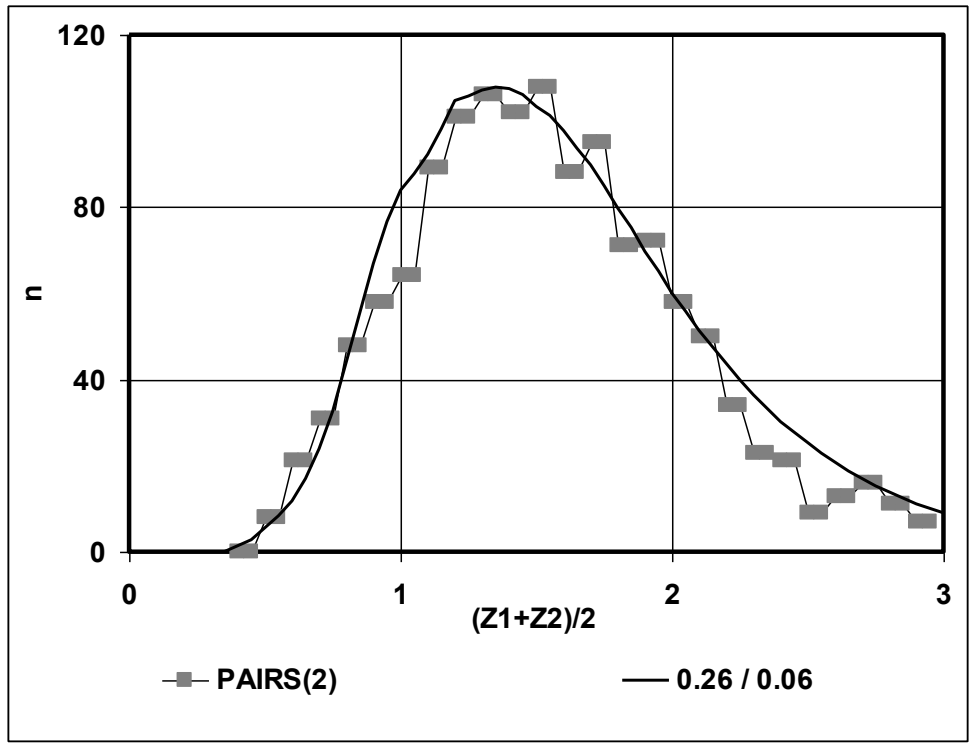

**Picture 24.** Calculated dependence n($Z_m$) for normal distribution $\beta_0$ with standard deviation 0.06 and average value 0.26 in comparison with the distribution of Picture 5.

This way the dependence of the amount of the pairs of opposite quasars from the value of the average red sift of the pair can be described exactly enough in the model of the closed Universe that expands with the relative velocity $\beta_0$ with average value 0.26 and standard deviation 0.06.

Getting back to the obtained earlier dependence (31) of the amount of observed objects from Z, let us study the possibility of its appliance to other existing data about the distribution of the quasars.

On the Picture 25 we show the distribution by Z objects that are classified by spectrum as quasars, from work [4]. On Picture 26 this distribution put to comparison with the dependence n(Z), calculated by the expression (31) for $\beta_0 = 0.22$ (chosen on basis of the best match of the disposition of the second maximum). At that the fact that draws the attention is an almost exact match of the dispositions of all the three maximums by the calculated and observed dependencies. But the correlation of the height of these maximums is significantly different.

Supposing that is can be related to the attenuation of the received signals in the space that should lead to the decreasing of the amount of the observed objects by means of their moving-off, let us enter to (31) an additional coefficient of the exponential descent exp(-kZ).

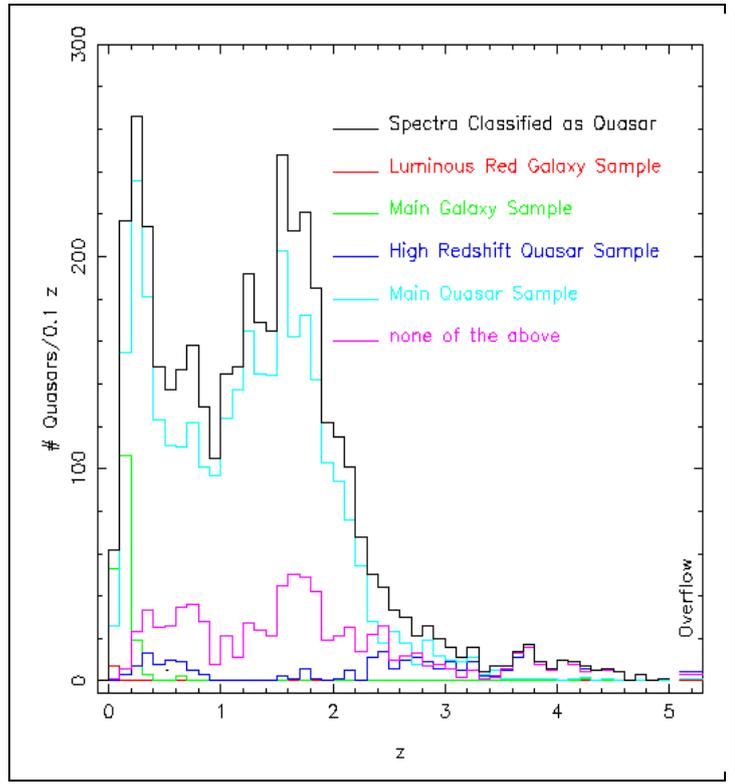

**Picture 25.** Distribution by Z objects classified by spectrum as quasars ([4], Figure 15: Redshift histogram for spectra classified as quasars).

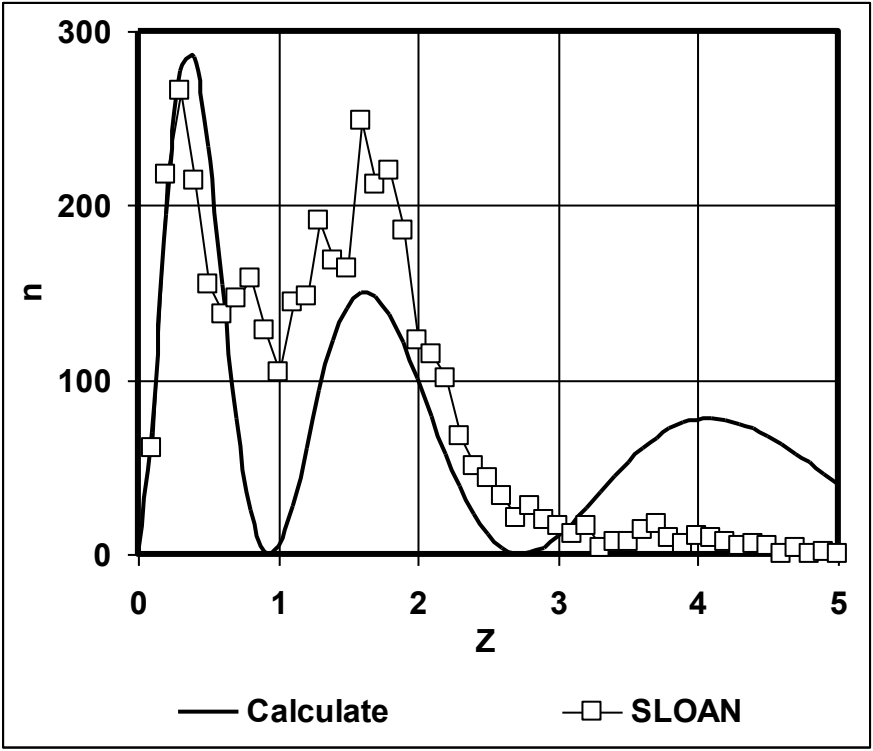

**Picture 26.** Distribution by Z objects classified by spectrum as quasars ([4], Figure 15: Redshift histogram for spectra classified as quasars) in comparison with the dependence n(Z), calculated by expression (31) for $\beta_0 = 0.22$.

$$n(Z) = (K/((Z+1)\beta_0)) \sin^2[\ln(Z+1)/\beta_0] \exp(-kZ); \qquad (42)$$

On the Picture 27 we show the distribution of Picture 25 in comparison with the dependence n(Z), calculated by the expression (42) for $\beta_0 = 0.22$ and $k = 0.8$ (selected on basis of the best match of the relation of the heights 2 and 3 of the maximums of the calculated and observed curves). At that the height of the first maximum of the calculated curve (not shown on the histogram) significantly exceeds the first maximum of the distribution of the observation data and makes $n \approx 1300$ at Z=0.3.

To explain possible reasons of such a mismatch let us turn to another distribution from the work [4] — distribution by Z of the objects classified by spectrum as galaxies, shown on Picture 28. On this curve the scale by Y-axis is 10 times bigger than on the curve of Picture 25, i.e. N/0.01z. The amount of the objects on 0.1z is from 3 to 40 thousands, and among them there is a component defined as quasars (Main Quasar Sample) with height 1-2 thousands of objects on 0.1z at 0<Z<0.3. On this basis we can suppose that the real curve of quasars distribution with consideration of the particularities of the classification can have a significantly higher than on Picture 25 peak at small values of Z, that has the value close to the first maximum of the calculated curve.

On Picture 29 we show the distribution of the Picture 25 in comparison with the dependence n(Z), calculated by the expression (42) for normal distribution $\beta_0$ with the average value 0.22 and standard deviation 0.03. The height of the first maximum of the calculated curve (not shown on the histogram) is $n \approx 1250$ at Z=0.3.

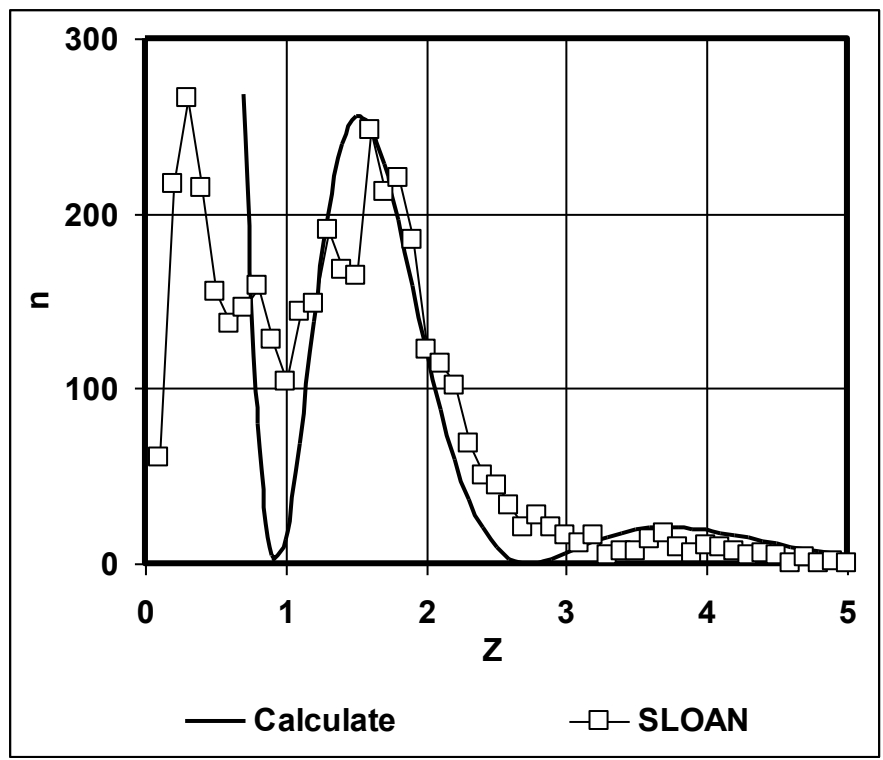

**Picture 27.** Distribution by Z objects classified by spectrum as quasars ([4], Figure 15: Redshift histogram for spectra classified as quasars) in comparison with the dependence n(Z), calculated by the expression (42) for $\beta_0 = 0.22$ и $k = 0.8$.

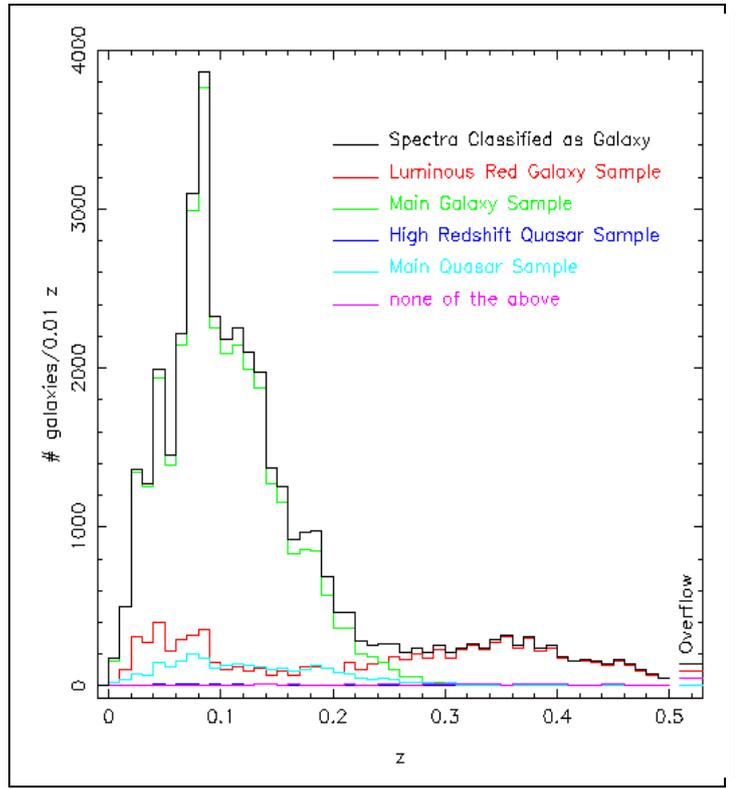

**Picture 28.** Distribution by Z objects classified by spectrum as galaxies ([A1], Figure 14: Redshift histogram for spectra classified as galaxies).

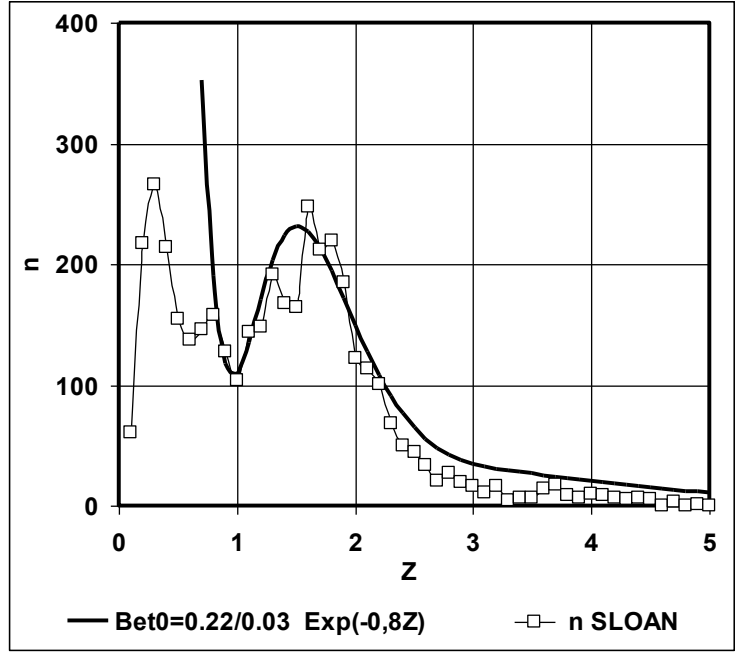

**Picture 29.** Distribution by Z objects classified by spectrum as quasars ([4], Figure 15: Redshift histogram for spectra classified as quasars) in comparison with the dependence n(Z), calculated by expression (42) for normal distribution $\beta_0$ with average value 0.22 and standard deviation 0.03.

Mentioned in [4] possibility to classify quasars as galaxies and galaxies as quasars can have the meaning for explanation of one of the reasons of small amount of discovered opposite quasars as it allows to suggest a hypothesis that at least some of the quasars can be the images of the galaxies that are located on a relatively small distances (Z<0.5), but came to us by the big circle ($\varphi_2$) and that is why appearing to be rather distant objects that have no strict structure because of the great way they made. In this case by the reason that the amount of galaxies is dozens times bigger than the amount of quasars, the objects that are centrally symmetrical to the most of the observed quasars can be galaxies or the objects classified as galaxies and that is why are absent in the quasars catalog [2].

The values of the relative velocity of the increasing of the radius of the Universe $\beta_0 \approx (0.2 - 0.25)$ that were used in our calculations, can be obtained in the standard cosmological model at very high values of the relative density of the matter.

$$\beta_0 = v/c = da/cdt = (\Omega-1)^{-1/2}. \qquad (43)$$

The values $\beta_0 < 0{,}25$ correspond to $\Omega > 17$, which absolutely contradicts to the observation data. However this contradiction can appear to be vincible. As it is known [3], the standard cosmological model is based on the metrics built up in neglect of the non-zero value of the differential of the scale factor in the non-stationary Universe. The attempt to consider this non-zero value can lead to the modification of the standard model [5]. In this case it appears that

$$\beta_0 = \alpha (1 - \alpha^2)^{1/2}; \qquad (44)$$

where

$$\alpha = \Omega^{1/2} (\Omega+1)^{-1/2}; \qquad (45)$$

from where

$$\beta_0 = \Omega^{1/2} / (\Omega+1); \qquad (46)$$

and at small values of $\beta_0$

$$\Omega \approx \beta_0^2 / (1 - \beta_0^2). \qquad (47)$$

By these values $\beta_0 \approx 0.2$ correspond to the values $\Omega \approx 0.05$, that are close to the observed values of the average density of the visible matter.

## 7. Conclusion

Performed in this work check of the supposition about the central symmetry of the celestial sphere and the possibility of its manifestation in the existence of centrally symmetrical pairs of the observed distant objects including the quasars brought up the existence of the pairs of centrally symmetrical (opposite) quasars in all the sections of the analyzed subequatorial zone of the celestial sphere.

The attempt to identify the discovered pairs of opposite quasars as two images of the same distant objects that reached the observer by two opposite arcs of the hypersphere circle has shown that in many cases the quasars that make a pair, has similar by form luminosity magnitudes profiles in the ranges **u,g,r,i,z** (from 300 to 1000 nm) with Pearson correlation coefficients close to 1 that can be considered as a proof of the possibility to identify them as two images of the same object.

According to the results of the statistical analysis the percentage of the pairs of the opposite quasars with correlation coefficients $R_{xy} > 0.98$ appears to be significantly higher than for the set of randomly formed pairs of quasars. Analysis of the dependence of this exceedance from the value of the artificial deviation of the central symmetry (Shift_RA) has shown that it practically disappears when one of the pair members is shifted to the direction of increasing or decreasing of the right ascension (RA) more than by 0.05 degree, which proves its regular, not random nature. According to the results of the additional check and statistical analysis, the probability of the random nature of the peaks does not exceed $10^{-9}$ (see Appendix).

This way, after the performed research we confirmed and statistically proved the phenomenon of the central symmetry of the celestial sphere that manifests in the existence of the centrally symmetrical pairs of quasars with identical individual characteristics (luminosity magnitudes profiles) that can be interpreted as images of the same object.

We have shown the possibility of the theoretical modeling of the observed dependencies of the amount of quasars and opposite pairs of quasars from the redshift in the closed Universe model. We supposed that the relatively small amount of the discovered in this work pairs of opposite quasars is related to the fact that the pair for most of the observed quasars can be galaxies or the objects classified as galaxies and that is why not included to the chosen as initial source of data quasar catalog SDSS-DR5.

**Acknowledgments**

We are very grateful for the invaluable opportunity to use the SDSS information.

Funding for the SDSS and SDSS-II has been provided by the Alfred P. Sloan Foundation, the Participating Institutions, the National Science Foundation, the U.S. Department of Energy, the National Aeronautics and Space Administration, the Japanese Monbukagakusho, the Max Planck Society, and the Higher Education Funding Council for England. The SDSS Web Site is http://www.sdss.org/.

The SDSS is managed by the Astrophysical Research Consortium for the Participating Institutions. The Participating Institutions are the American Museum of Natural History, Astrophysical Institute Potsdam, University of Basel, University of Cambridge, Case Western Reserve University, University of Chicago, Drexel University, Fermilab, the Institute for Advanced Study, the Japan Participation Group, Johns Hopkins University, the Joint Institute for Nuclear Astrophysics, the Kavli Institute for Particle Astrophysics and Cosmology, the Korean Scientist Group, the Chinese Academy of Sciences (LAMOST), Los Alamos National Laboratory, the Max-Planck-Institute for Astronomy (MPIA), the Max-Planck-Institute for Astrophysics (MPA), New Mexico State University, Ohio State University, University of Pittsburgh, University of Portsmouth, Princeton University, the United States Naval Observatory, and the University of Washington.

SDSS Technical Paper References: [6-12]. We also thank Dr. Feofilov for the friendly support in the preparation of this work to the submission.

# APPENDIX

Re-check the data on the dependence of the number of pairs of opposite quasars with a high degree of correlation of the luminosity profiles ($R_{xy} \to 1$) of disorders of the central symmetry, given the introduction of the shift (Shift_RA) to the coordinates quasars located at $180^0 < RA < 360^0$. The dependences obtained are shown in Table A1.

| № п/п | SHIFT_RA (degree) | N_pairs (dZ<0.3) | | | | N_pairs (dZ<0.2) | | | |
|---|---|---|---|---|---|---|---|---|---|
| | | ALL pairs | Rxy>0.97 | Rxy>0.98 | Rxy>0.99 | ALL pairs | Rxy>0.97 | Rxy>0.98 | Rxy>0.99 |
| 1. | -0,500 | 286 | 77 | 51 | 24 | 186 | 64 | 44 | 19 |
| 2. | -0,450 | 295 | 77 | 50 | 27 | 197 | 62 | 42 | 22 |
| 3. | -0,400 | 328 | 72 | 48 | 28 | 214 | 61 | 42 | 25 |
| 4. | -0,350 | 343 | 89 | 56 | 32 | 231 | 73 | 48 | 29 |
| 5. | -0,300 | 324 | 87 | 56 | 27 | 217 | 70 | 45 | 23 |
| 6. | -0,250 | 314 | 79 | 49 | 22 | 201 | 63 | 38 | 19 |
| 7. | -0,20 | 318 | 90 | 55 | 28 | 212 | 72 | 42 | 22 |
| 8. | -0,175 | 301 | 91 | 58 | 31 | 194 | 70 | 42 | 24 |
| 9. | -0,15 | 307 | 94 | 62 | 30 | 196 | 74 | 47 | 24 |
| 10. | -0,125 | 311 | 96 | 65 | 32 | 203 | 77 | 49 | 27 |
| 11. | -0,10 | 310 | 90 | 61 | 28 | 207 | 71 | 46 | 25 |
| 12. | -0,075 | 316 | 102 | 71 | 31 | 212 | 82 | 54 | 28 |
| 13. | -0,05 | 320 | 109 | 80 | 33 | 212 | 85 | 61 | 28 |
| 14. | -0,025 | 319 | 99 | 71 | 30 | 213 | 76 | 54 | 25 |
| 15. | 0 | **329** | **103** | **79** | **38** | **214** | **81** | **64** | **32** |
| 16. | 0,025 | 346 | 106 | 82 | 38 | 228 | 85 | 68 | 31 |
| 17. | 0,05 | 318 | 89 | 63 | 29 | 209 | 71 | 52 | 23 |
| 18. | 0,075 | 302 | 89 | 61 | 29 | 208 | 72 | 53 | 24 |
| 19. | 0,10 | 303 | 81 | 54 | 27 | 205 | 64 | 46 | 22 |
| 20. | 0,125 | 304 | 74 | 48 | 26 | 200 | 59 | 43 | 22 |
| 21. | 0,15 | 310 | 75 | 47 | 23 | 204 | 60 | 41 | 19 |
| 22. | 0,175 | 319 | 76 | 50 | 22 | 217 | 60 | 44 | 19 |
| 23. | 0,20 | 308 | 79 | 61 | 27 | 215 | 64 | 52 | 22 |
| 24. | 0,250 | 308 | 78 | 54 | 23 | 211 | 60 | 44 | 17 |
| 25. | 0,300 | 300 | 88 | 59 | 30 | 186 | 65 | 45 | 21 |
| 26. | 0,350 | 328 | 91 | 59 | 30 | 213 | 68 | 46 | 23 |
| 27. | 0,400 | 305 | 89 | 65 | 35 | 206 | 67 | 50 | 27 |
| 28. | 0,450 | 325 | 85 | 63 | 33 | 218 | 65 | 49 | 26 |
| 29. | 0,500 | 326 | 86 | 62 | 32 | 213 | 64 | 45 | 25 |
| | MEAN: | 315 | | | | 210 | | | |

**Table A1.** Numbers of opposing pairs of quasars that match the criteria (dZ;Rxy) for different Shift_RA.

The above data are compared with the results of the statistical analysis of the likelihood of pairs of random deviation from the mean values corresponding to the array before performing the assay 40,000 randomly drawn pairs, according to the formula of the binomial distribution. If there are N objects (pairs of quasars), which are known probability p of being in a given state (Rxy>0.98), then the probability P(n) the fact that in this state there are n objects from N, is given by

$$P(n) = p^n q^{N-n} N!/n!(N-n)!; \qquad (A1)$$

where $q \equiv 1-p$.

Probability matching condition (Rxy>0.98) for an array of 40,000 random pairs of quasars for dZ<0.3 is 0.1777 (17.77%), the average sample size N ≈ 315. According to the formula (A1) that the number of pairs corresponding to different levels of risk of accidental deviations from the mean, as shown in Table 2A and Picture 1A.

| The criteria (dZ;Rxy) | $N_{mean}$ | $n_{mean}$ | p<0.1 | | p<0.01 | p<0.001 | p<0.0001 |
|---|---|---|---|---|---|---|---|
| (dZ<0.2, Rxy>0,97) | 210 | n≈64 | n<56 | n>73 | n>81 | n>85 | n>90 |
| (dZ<0.2, Rxy>0,98) | 210 | n≈44 | n<37 | n>52 | n>58,5 | n>64,5 | n>68,5 |
| (dZ<0.2, Rxy>0,99) | 210 | n≈21,5 | n<16 | n>27 | n>32 | n>36 | n>39 |
| (dZ<0.3, Rxy>0,97) | 315 | n≈83 | n<73 | n>93 | n>102 | n>108 | n>113 |
| (dZ<0.3, Rxy>0,98) | 315 | n≈56,5 | n<48 | n>65 | n>73 | n>78 | n>83 |
| (dZ<0.3, Rxy>0,99) | 315 | n≈26 | n<20 | n>33 | n>38 | n>43,5 | n>46 |

**Table 2A.** The results of the calculation of the probability characteristics by the formula of the binomial distribution.

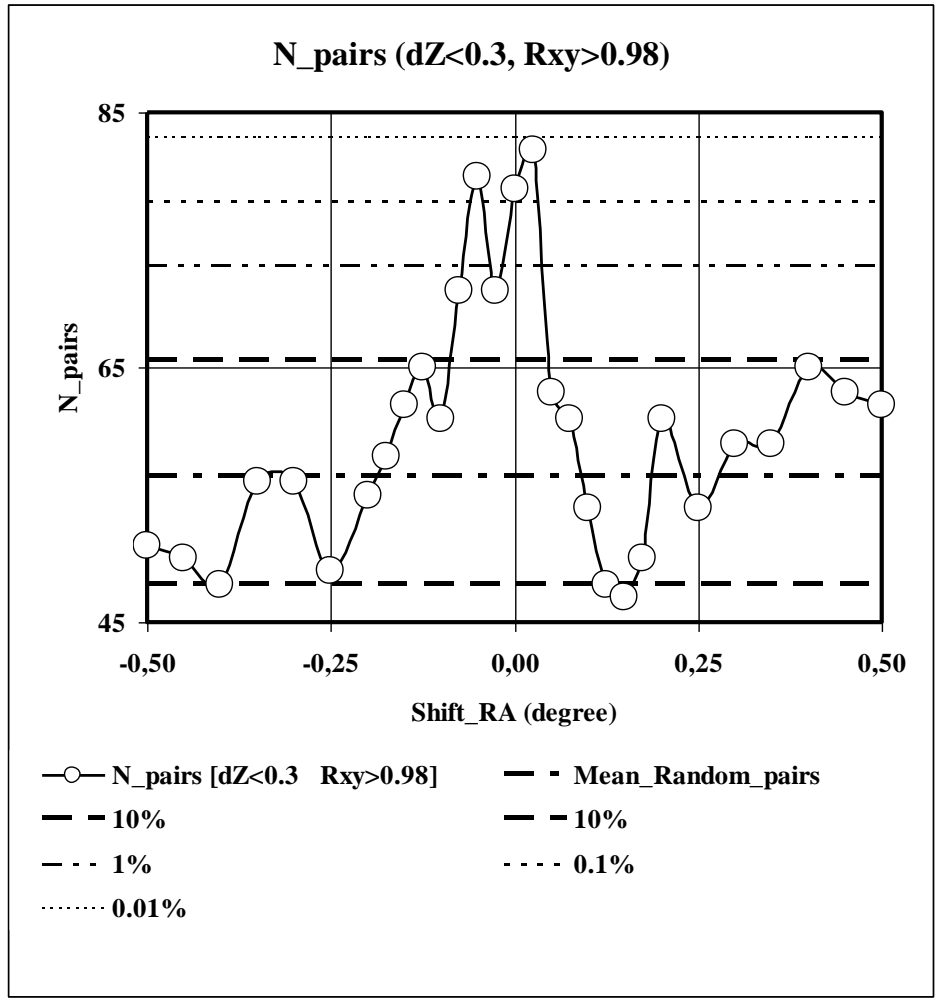

**Picture 1A.** The dependence of the number of pairs of opposed quasars that match the criteria dZ<0,3;Rxy>0,98 on Shift_RA. The horizontal lines show the number of pairs corresponding to the average volume of the sample (N ≈ 315) and can match the specified criteria (p = 0.1777) for an array of randomly drawn pairs. The dot-dashed line - the average amount for a sample of this size, the dotted line - the probability of a random deviation from the mean.

Our results demonstrate the presence of a maximum in the range ($-0.075^0 <$ Shift_RA $< 0.025^0$). The maximum includes a 3 points, for which the probability of a random deviation p1, p2, p3 are in the ranges: $10^{-3} > p1 > 10^{-4}$, $10^{-3} > p2 > 10^{-4}$, $10^{-3} > p3 > 10^{-4}$. Thus, the probability of its formation due to random variation points from the average does not exceed $P_{max} = p1p2p3 < 10^{-9}$, which eliminates the possibility of a random nature and confirms the existence of opposing pairs of quasars with identical characteristics of luminosity, which may be a pair of images of the same object.